\documentclass[journal]{IEEEtran}
\usepackage{eurosym}
\usepackage{microtype}
\usepackage{amssymb}
\usepackage{yfonts}
\usepackage{cite}
\usepackage{url}
\usepackage{multirow}
\usepackage{xcolor}
\usepackage{color}
\usepackage{blindtext}
\usepackage{eucal}
\usepackage{relsize}
\usepackage{algorithmic}
\usepackage{tikz}
\newcommand*\circled[1]{\tikz[baseline=(char.base)]{
            \node[shape=circle,draw,inner sep=1.5pt] (char) {#1};}}

\usepackage{nomencl}
\usepackage{etoolbox}
\makenomenclature

\renewcommand\nomgroup[1]{%
\color{black}
  \item[\bfseries
  \ifstrequal{#1}{P}{Variables and functions}{%
  \ifstrequal{#1}{N}{Constants}{%
  \ifstrequal{#1}{O}{Sets and Indices}{}}}%
]\vspace{5pt}}

\usepackage[cmex10]{amsmath}

\usepackage{amsfonts}

\usepackage{array}

\usepackage[caption=false]{subfig}

\makeatletter
\newcommand{\thickhline}{%
    \noalign {\ifnum 0=`}\fi \hrule height 0.75pt
    \futurelet \reserved@a \@xhline
}
\newcolumntype{"}{@{\hskip\tabcolsep\vrule width 0.75pt\hskip\tabcolsep}}
\makeatother

\begin{document}
%
%
%
%

\title{Fast Resource Scheduling for Distribution Systems Enabled with Discrete Control Devices}

\author{Alireza~Nouri,~\IEEEmembership{Member,~IEEE,},
Alireza~Soroudi,~\IEEEmembership{Senior~Member,~IEEE,}
and~Andrew~Keane,~\IEEEmembership{Senior~Member,~IEEE}
\thanks{
A. Nouri (alireza.nouri@ucd.ie), A. Soroudi (alireza.soroudi@ucd.ie) and A. Keane (andrew.keane@ucd.ie) are with the School of Electrical and Electronic Engineering, University College Dublin, Dublin 04, Ireland.

This work has emanated from research conducted with the financial support of Science Foundation Ireland under the SFI Strategic Partnership Programme Grant Number SFI/15/SPP/E3125. The opinions, findings and conclusions or recommendations expressed in this material are those of the authors and do not necessarily reflect the views of the Science Foundation Ireland.}
}

\markboth{}%
{Shell \MakeLowercase{\textit{et al.}}: Bare Demo of IEEEtran.cls for Journals}
\maketitle

\begin{abstract}
This paper proposes a framework for fast short-term scheduling and steady state voltage control in distribution systems enabled with both continuous control devices, e.g., inverter interfaced DGs and discrete control devices (\textsc{dcd}s), e.g., on-load tap changers (\textsc{oltc}s). The voltage-dependent nature of loads is taken into account to further reduce the operating cost by managing the voltage levels. 
Branch and cut method is applied to handle the integrality constraints associated with the operation of \textsc{dcd}s. A globally convergent trust region algorithm (\textsc{tra}) is applied to solve the integer relaxed problems at each node during the branching process. The \textsc{tra} sub-problems are solved using interior point method.
To reduce the branching burden of branch and cut algorithm,
before applying \textsc{tra} at each node, a simplified optimization problem is first solved.
Using the convergence status and value of objective function of this problem, a faster decision is made on stopping the regarding branch. Solving the simplified problem obviates the application of \textsc{tra} at most nodes.
It is shown that the method converges to the optimal solution with a considerable saving in computation time according to the numerical studies. 
\end{abstract}
\IEEEpeerreviewmaketitle


\maketitle

\nomenclature[P, 01]{$r_t$}{Turn ratio of transformer $t$ (pu)}
\nomenclature[P, 02]{$st$}{Step of each \textsc{cb} which is inserted into the circuit}
\nomenclature[P, 03]{$tap_t$}{Tap position of \textsc{oltc} $t$}
\nomenclature[P, 05]{$I_l$}{Current of line $l$}
\nomenclature[P, 07]{$I^*$}{Conjugate of vector $I$}
\nomenclature[P, 07]{$N_p$}{Number of transformer primary turn}
\nomenclature[P, 08]{$P_c$}{Transformer core loss}
\nomenclature[P, 09]{${P_d}_b$}{Voltage dependent active power demand at bus $b$}
\nomenclature[P, 10]{$P_g$/$Q_g$}{Active/reactive power generated by continuous \textsc{cd}s}
\nomenclature[P, 11]{$P_p$/$Q_p$}{Active/reactive power injected by upstream network}
\nomenclature[P, 12]{${Q_d}_b$}{Voltage dependent reactive power demand at bus $b$}
\nomenclature[P, 16]{$V_b$}{Voltage phasor at bus $b$}
\nomenclature[P, 17]{$\left|V_b\right|$}{Magnitude of voltage at bus $b$}
\nomenclature[P, 18]{$V_{th}$}{Upstream system Thevenin voltage}
\nomenclature[P, 19]{$W$}{Vector of control variables, e.g., $Pg$}
\nomenclature[P, 20]{$Z_{pr,p}$}{Parallel impedance at the primary side of $\pi$ model}
\nomenclature[P, 21]{$Z_{pr,s}$}{Parallel impedance at the secondary side of $\pi$ model}
\nomenclature[P, 22]{$Z_{sr}$}{Series impedance in the equivalent $\pi$ model}
\nomenclature[P, 23]{$\delta_V$}{Angle of phasor V}
\nomenclature[P, 24]{$\Delta \Psi$}{Vector of perturbed optimization variables, i.e., $\Psi$-$\hat{\Psi}$}
\nomenclature[P, 26]{$\epsilon^+_b$/$\epsilon^{-}_b$}{Slack variable of upper/lower voltage limit at bus $b$}
\nomenclature[P, 27]{$\varphi_{I,V}$}{Phase difference between $I$ and $V$}
\nomenclature[P, 28]{$\Psi$}{Vector of optimization variables, e.g., $Pg$ and $V_b$}
\nomenclature[P, 29]{$\psi_x$/$\psi_y$}{Real/imaginary part of phasor $\phi$}

\nomenclature[N, 01]{$y^{st}$}{\textsc{cb} admittance step change}
\nomenclature[N, 02]{$P_{d_0}$}{Active power demand at voltage $V_0$}
\nomenclature[N, 03]{$Q_{d_0}$}{Reactive power demand at voltage $V_0$}
\nomenclature[N, 04]{$R_c^n$}{Nominal transformer core resistance (pu)}
\nomenclature[N, 05]{$S^{base}$}{Base value for power}
\nomenclature[N, 06]{$S^n$}{Capacity of continuous control devices}
\nomenclature[N, 11]{$X_M^n$}{Nominal transformer magnetizing reactance (pu)}
\nomenclature[N, 12]{$Y_\text{Bus}$}{Network admittance matrix}
\nomenclature[N, 13]{${Y_l}$}{Admittance of line $l$, i.e., 1/${X_l}$}
\nomenclature[N, 14]{$Z_p$/$Z_s$}{Transformer primary/secondary series impedance (pu)}
\nomenclature[N, 16]{$Z_t^n$}{Nominal series impedance of transformer $t$ (pu)}
\nomenclature[N, 17]{$Z_{th}$}{Upstream system Thevenin impedance}
\nomenclature[N, 18]{$\alpha^{\text{max},pv}$}{Maximum power angle of photovoltaic units}
\nomenclature[N, 19]{${\Delta}U$}{Voltage step change per tap in pu}
\nomenclature[N, 20]{${\kappa_p}$/${\kappa'_p}$}{Constant power share in ZIP/ZP load model}
\nomenclature[N, 22]{${\mu_p}$}{Constant current share in ZIP load model}
\nomenclature[N, 23]{${\zeta_p}$/${\zeta'_p}$}{Constant impedance share in ZIP/ZP load model}
\nomenclature[N, 25]{$\rho_A$/$\rho_R$}{Active/reactive power consumption price}
\nomenclature[N, 26]{$\rho_i$}{Energy price of \textsc{der} $i$ (\euro/Wh)}
\nomenclature[N, 28]{$\tau$}{Scheduling time period in hour}
\nomenclature[N, 29]{$\hat{\Psi}$}{Vector of optimization variables $\Psi$ found in the previous \textsc{tra} sub-problem or vector of initial solution.}
\nomenclature[O, 04]{$rr$}{Set of \textsc{rr}s}
\nomenclature[O, 07]{$vc$}{Superscripts to indicate voltage control mode}
\nomenclature[O, 08]{$N_\text{B}$}{Number of of buses indexed by $b$}
\nomenclature[O, 10]{$N_s$}{Number of transformer secondary turn}
\nomenclature[O, 11]{$N_t$}{Number of parallel transformers indexed by $t$}
\nomenclature[O, 12]{$N_\text{W}$}{Number of control variables}
\printnomenclature
\section*{\textcolor[rgb]{0,0,0}{List of Abbreviations}}
\addcontentsline{toc}{section}{List of Abbreviations}
\begin{IEEEdescription}[\IEEEusemathlabelsep\IEEEsetlabelwidth{AAAa}]
\color{black}
\item[\textsc{bc}]    Branch and cut
\item[\textsc{cb}]    Capacitor bank
\item[\textsc{dcd}]   Discrete controllable device
\item[\textsc{der}]   Distributed energy resource
\item[\textsc{dsp}]   Distribution scheduling problem
\item[\textsc{fr}]    Feasible region
\item[\textsc{minlp}] Mixed integer nonlinear programming
\item[\textsc{nlp}]   Nonlinear programming
\item[\textsc{oltc}]  On load tap changer
\item[\textsc{opf}]   Optimal power flow
\item[\textsc{rr}]    Renewable resource
\item[\textsc{sdp}]   Semidefinite programming
\item[\textsc{socp}]  Second-order cone programming
\item[\textsc{svr}]   Static voltage regulator
\item[\textsc{tra}]   Trust region algorithm
\end{IEEEdescription}
\section{Introduction}\label{sec_1}
\IEEEPARstart{T}{he resource} scheduling in current distribution systems is a critical task due to the high penetration of distributed energy resources (\textsc{der}s), fast load variations \cite{Watson2018}, presence of several controllable devices, \textcolor[rgb]{0,0,0}{possible conflict between the operation of different controllable devices}, and other challenges \cite{KIANMEHR2019471}.
\textcolor[rgb]{0,0,0}{This paper proposes a novel technique to find the globally optimal solution of the distribution resource scheduling problem. These resources include \textsc{der}s, static voltage regulators (\textsc{svr}s), renewable resource (\textsc{rr}s) as the fast continuous controllable devices and capacitor banks (\textsc{cb}s) and on load tap changers (\textsc{oltc}s) as the slow discrete control devices (\textsc{dcd}s).
A fair share of efforts is focused on expediting the solution process to meet the near-real-time requirements of the intended application.}
The distribution scheduling problem (\textsc{dsp}) is a mixed integer nonlinear programming (\textsc{minlp}) problem that takes a long time to converge to a solution with no guarantee on solution optimality.
\textcolor[rgb]{0,0,0}{With slow optimization techniques, the solution might not be optimal in the time of application due to the change of problem input parameters, e.g., load levels. A fast technique is required to obtain the solution which better complies with the system state. However, the solution accuracy cannot be jeopardized.}

Branch and cut (\textsc{bc}) technique is adopted here to deal with the integer and binary variables.
\textcolor[rgb]{0,0,0}{At every node during the branching process, an integer relaxed problem (in which all the variables are assumed to be continuous) is solved. \textsc{bc} technique assumes that an algorithm, called ``sub-algorithm'', exists to solve the continuous problems with the additional bounds on integer variables.
A globally convergent \cite{Songqiang2018} trust region algorithm (\textsc{tra}) \cite{Sheng2014} is applied here as the sub-algorithm.
\textsc{tra}s are iterative algorithms for solving nonlinear optimization problems. In every iteration, \textsc{tra}s solve a ``sub-problem'' to minimize a quadratic approximation (model function) of the nonlinear objective function in a restricted vicinity (trust region) of the initial guess or the solution point obtained in the previous sub-problem. 
After solving each sub-problem, the reduction in the model function should be equal to the reduction in the nonlinear objective function within an acceptable tolerance.}
If this is not the case, the trust region is contracted and the approximated model is solved again.

The constraint on the step size and problem original constraints may be inconsistent. This might render most sub-problems infeasible. Byrd-Omojokun technique \cite{Songqiang2018} is applied to cope with this inconsistency.
\textcolor[rgb]{0,0,0}{This technique decomposes each sub-problem into two sub-problems which are easier to solve. This makes the technique more favorable for large-scale problems.
This ability of \textsc{tra}s to deal with inconsistent constraints makes them globally convergent algorithms, compared to the other sequential optimization approaches.
Interior point method is used to solve the decomposed sub-problems.
After solving each sub-problem, a power flow (\textsc{pf}) algorithm is applied to calculate the exact voltages.} Such voltages are used to update the model function for the next sub-problem.

\textcolor[rgb]{0,0,0}{This solution technique is efficient in terms of optimality based on the numerical results. It is also fast enough when applied to solve the \textsc{dsp} for medium-scale systems. However, it is necessary to improve the convergence speed to meet the near real-time requirements.
To this end, before applying \textsc{tra} at each node during the branching process, a simplified problem is first solved.
Based on the solution of this problem, solving the \textsc{tra} sub-problem is avoided for \textit{most} nodes.
The simplified problem is, in fact, a linear programming (\textsc{lp}) problem.
Based on the discussion provided in Section \ref{SecIIIC},
if at a certain node, the solution of \textsc{lp} problem is dominated by the best integer-feasible solution seen so far or is infeasible, the accurate solution is also dominated or infeasible and the current branch is not to be further continued. This obviates application of \textsc{tra} for such nodes and leads to a huge saving in computation time.}

The voltage-dependent nature of loads and accurate models of \textsc{oltc} transformers are taken into account to further reduce the operating cost by managing the voltage levels. The parameters of upstream system and load models are kept up-to-date  while solving the \textsc{dsp}. 
\subsection{Review of the Related Literature}\label{LitRev}
\textcolor[rgb]{0,0,0}{The ac optimal \textsc{pf} (\textsc{opf}) problem was transformed into a non-iterative convex problem in \cite{Watson2018} in the absence of \textsc{dcd}s.
The type of all loads was considered to be constant current and it was assumed that controllable devices can be modeled as independent current injections. Considering these assumptions the method was proved to be fast enough when applied on a distribution system enabled with energy storage systems. The first assumption restricts the type of loads and the second assumption implies small voltage deviations which is not the case with most of the practical distribution systems.
Compared to \cite{Watson2018}, in this paper, no restricting assumptions are made, \textsc{dcd}s are included in the model and the upstream system and load models are kept updated while solving the \textsc{dsp}.}

\textcolor[rgb]{0,0,0}{An iterative optimization was used in \cite{Dao2017} to solve the \textsc{dsp}.
Similar to the present paper, the voltage-dependent load model and accurate model of \textsc{oltc} transformers were included.
The problem was converted to an iterative least square optimization with linear constraints. Though the solution quality was shown to be high enough in the case studies, there is no guarantee for global optimality.
In the case that tap positions were not integer, they were rounded to the nearest integer values. This may render the solution suboptimal and even infeasible.}

\textcolor[rgb]{0,0,0}{Sequential quadratic \textsc{tra} was applied in \cite{Sheng2014} to simultaneously minimize the energy loss and voltage deviation and maximize the production of \textsc{der}s. The multi-objective problem was converted to a single objective problem through normalization.
In the present paper, the cost of both copper and iron losses are included the objective function, the voltage deviation problem is modeled as soft constraints and the generations of \textsc{der}s are optimized along with the other controllable parameters. A \textsc{minlp} technique is proposed to handle the \textsc{dcd}s and an expediting mechanism is also proposed. The mathematical background and details of applying \textsc{tra} to solve constrained \textsc{nlp} problems were presented in \cite{BAHRAMI2020}.}
%

\textcolor[rgb]{0,0,0}{Semidefinite programming (\textsc{sdp}) techniques, especially second order cone programming
(\textsc{socp}) based on branch flow model
are able to solve the \textsc{dsp} in balanced distribution systems
enabled with continuous control devices under quite acceptable assumptions.
The formulations based on conic relaxation enable application of commercial solvers and therefore, are able to reduce the solution time. However, the simplifying assumptions restrict the application of these techniques.
\textsc{socp} was applied in \cite{Stoch} to solve the coordinated optimization of active and reactive powers in balanced systems.
The optimal active and reactive power dispatch was found for a long time period, e.g., 24 hours, to cope with the uncertainties.
The transformer primary voltage was assumed to be independent of the control variables to extract a model for \textsc{oltc} transformers that keeps the conic convexity.
The effects of voltage level on \textsc{cb}s' reactive power injection were also neglected for the same reason. Here, the exact models are presented for \textsc{oltc}s and \textsc{cb}s with no restricting assumptions. Since the main focus of this paper is on the solution technique, the uncertainties are neglected. However, they can be included following the same two-stage method proposed in \cite{Stoch}.
A model was developed in \cite{LinearOLTC} for \textsc{dcd}s to obviate the restricting assumption of fixed primary voltages in branch flow model by introducing auxiliary binary variables.
These binary variables drastically increase the solution time.}

\textcolor[rgb]{0,0,0}{For unbalanced systems, due to mutual inductances and unbalanced currents and voltages, the \textsc{dsp} cannot be reformulated as an \textsc{sdp} problem based on branch flow model.
A formulation was proposed in \cite{Unbalanced} to include \textsc{oltc} transformers in branch flow model neglecting the iron losses and effects of tap-changing operations on transformer series impedance.
\textsc{sdp} was applied in \cite{Unbalanced} and \cite{UnbalancedAlaki} to solve the \textsc{dsp} in unbalanced systems neglecting the voltage unbalances and mutual inductances, respectively.
Such assumptions are not acceptable for practical systems, where the voltage unbalance really matters.
Unlike these methods, the proposed method and expediting approach can be applied on unbalanced distribution systems.}

It is imperative to use an accurate model for \textsc{oltc} transformers to show the effect of tap-changing operations on the copper and iron losses as well as the system demand.
An accurate and adaptive voltage-dependent load model is also required \cite{Arefifar2013}. \textcolor[rgb]{0,0,0}{A method for updating the load model parameters was proposed and tested on a real-life system in \cite{Arefifar2013}.} This method is modified here to keep the load models updated while solving the \textsc{dsp}.
The accuracy of the upstream system model is another important factor.
\textcolor[rgb]{0,0,0}{It was shown in \cite{Bahadornejad2014} that the changes in the voltage and current at the primary side of transformer can be used to estimate the upstream system Thevenin impedance and to monitor the \textsc{oltc} stability.}
This method is modified and used here to keep both Thevenin impedance and voltage updated according to the solution optimality concerns.
\subsection{List of Contributions}\label{Contrib}
\begin{enumerate}
\item to propose a globally convergent solution methodology for resource scheduling in the presence of \textsc{dcd}s with no simplifying assumptions.
\item to expedite the solution by reducing the branching burden based on the solution of a simplified problem before engaging in solving the nonlinear \textsc{opf} at each node.
\item to keep the loads and upstream system models up-to-date while solving the \textsc{dsp}.
\end{enumerate}
\section{Scheduling Framework}\label{SFrame}
\begin{figure}[!t]
  \centering
  \includegraphics[width=0.86\columnwidth,clip=true]{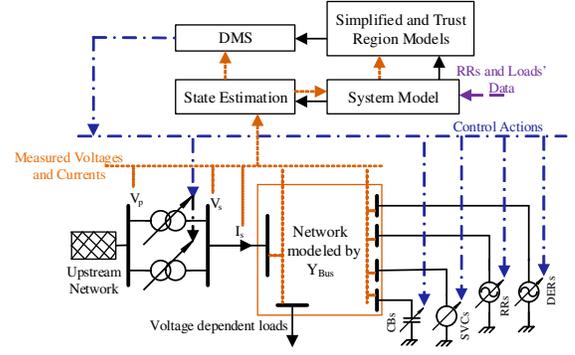}
	\caption{Distribution system and scheduling framework.}\label{System} \vspace{-3mm}
\end{figure}

Fig. \ref{System} shows the controllable devices working under a central control scheme.
The solid lines show how to develop the models, dashed lines show the flow of the measured/estimated data and dashed-dotted lines show the control commands. The measured or estimated voltages and currents are used to develop the simplified and trust region optimization formulations.

The discrete variables include the independent per-phase tap positions, steps of the \textsc{cb}s and auxiliary binary variables. For uncertainty handling, a longer horizon is considered \cite{Ren2018}. The longer the scheduling horizon, the higher the number of integer variables.
Here, the main focus is on the solution technique. Therefore, to keep the narrative simple, some complicating aspects are neglected bearing in mind that the proposed technique should be able to solve the problem with all these aspects considered. For instance, it is assumed that the system is balanced. The details of applying \textsc{tra} for solving unbalanced \textsc{opf} is found in \cite{Li2016}.

To handle the integer variables, a \textsc{bc} technique is implemented in MATLAB.
To find the optimal solution of the integer relaxed \textsc{nlp} problems during the branching process, \textsc{tra} is applied. Interior point technique is used to solve the \textsc{tra} sub-problems using IPOPT in GAMS. The next focus is on expediting the solution based on the results of solving an \textsc{lp} problem to obviate application of \textsc{tra} at most nodes (Section \ref{SecIIIC}). The \textsc{lp} problems are solved using CPLEX in GAMS.
\subsection{Application of TRA to Solve the Integer-Relaxed Problems}\label{TRA0}
In each \textsc{tra} sub-problem, interior point method minimizes a \textit{quadratic} approximation of the objective function, as the model function, subjected to the linearized constraints, within a trust region around the candidate solution, i.e., $\hat{\Psi}$. For continuous control devices, control variables ($W$) include $P_g$ and $Q_g$ under power control mode and $P_g$ and terminal voltages under voltage control mode. For \textsc{dcd}s, $W$ includes $tap$ and $st$. The objective is minimization of the total system cost \eqref{OBJ}.
\begin{equation}
\label{OBJ}
\begin{array}{l}
\displaystyle
\hspace{-4mm}
\underbrace{ Min }_{W} \left\{F\right\} \hspace{-1mm}
= \hat{F}
+\tau S^{base}(\rho_A\Delta P_p \hspace{-1mm}
+ \hspace{-1mm}
\rho_R\Delta Q_p \hspace{-1mm}
+ \hspace{-2mm}
\sum_{i\in{der}}{\hspace{-1mm} \rho_i{\Delta P_g}_i})
\end{array}
\hspace{-3mm}
\end{equation}

To find the model function, $\Delta P_p$ and $\Delta Q_p$ in \eqref{OBJ} should be approximated by two quadratic functions in terms of $W$.
A model is presented for \textsc{oltc} transformer(s) in \ref{SecA}. It is used to map $\Delta V_s$ and $\Delta I_s$ to $\Delta V_p$ and $\Delta I_p$ \eqref{Vs2Vp}. The entries of matrix $T$ depend on the tap positions and will be introduced in subsection \ref{SecA}. \textcolor[rgb]{0,0,0}{The perturbed form of \eqref{Vs2Vp} is provided in \eqref{Vs2VpPer}. The relationships between $\Delta P_p$ and $\Delta Q_p$ and $\Delta V_p$ and $\Delta I_p$ is given in \eqref{PinQin0}. Replacing \eqref{Vs2VpPer} in \eqref{PinQin0}, the relationships between $\Delta P_p$ and $\Delta Q_p$ and $\Delta V_s$, $\Delta I_s$ and $\Delta tap$ are found.}

A perturbed model will be developed in subsection \ref{SecIII} to give $\Delta V_s$ and $\Delta I_s$ in terms of $\Delta W$. Using this model, \eqref{Vs2VpPer} and \eqref{PinQin0}, $\Delta P_p$ and $\Delta Q_p$ are expressed as quadratic functions of $\Delta W$ (vector $tap$ is also included in $W$).
%
\begin{equation}
\label{Vs2Vp}
\begin{array}{l}
\displaystyle
\begin{pmatrix} {V}_p \\ {I}_p \end{pmatrix}
=
\underbrace{\begin{pmatrix} 1+\frac{Z_{sr}}{Z_s}  &  Z_{sr}
\vspace{1mm}
\\
\frac{1}{Z_s}+\frac{1}{Z_p}+\frac{Z_{sr}}{Z_sZ_p}  & 1+\frac{Z_{sr}}{Z_p} \end{pmatrix}}_{T(tap)} 

\begin{pmatrix} {V}_s \\ {I}_s \end{pmatrix} 
\end{array}
\end{equation}
\begin{equation}
\label{Vs2VpPer}
\begin{array}{l}
\displaystyle
\hspace{-4.5mm}
\begin{pmatrix} \Delta{V}_p \\ \Delta{I}_p \end{pmatrix} \hspace{-1.25mm}
= \hspace{-1.25mm}
\begin{pmatrix} \frac{\partial T_{11}}{\partial tap} \Delta tap & \frac{\partial T_{12}}{\partial tap} \Delta tap
\vspace{1mm}
\\
\frac{\partial T_{21}}{\partial tap} \Delta tap & \frac{\partial T_{22}}{\partial tap} \Delta tap \end{pmatrix}
\hspace{-1mm}
\begin{pmatrix} \hat{V}_s \\ \hat{I}_s \end{pmatrix}
\hspace{-1mm} + \hspace{-0.5mm} T(tap)
 \hspace{-1mm}
\begin{pmatrix} \Delta{V}_s \\ \Delta{I}_s \end{pmatrix} \hspace{-1mm}
\end{array}
\hspace{-3mm}
\end{equation}
\begin{equation}
\label{PinQin0}
\begin{array}{l}
\displaystyle
\hspace{-4mm}
\begin{pmatrix} \Delta{P}_p \\ \Delta{Q}_p \end{pmatrix} \hspace{-1mm}
= \hspace{-1.25mm}
\begin{pmatrix} {\hat{V_p}}_x \hspace{-3mm} & \hspace{-2mm} {\hat{V_p}}_y\\ {\hat{V_p}}_y \hspace{-3mm} & \hspace{-2mm} -{\hat{V_p}}_x \end{pmatrix} \hspace{-0.75mm}
 \hspace{-1mm}
\begin{pmatrix} \Delta{I_p}_x \\ \Delta{I_p}_y \end{pmatrix} \hspace{-1.25mm}
\vspace{1mm}
+ \hspace{-1.25mm}
\begin{pmatrix} {\hat{I_p}}_x \hspace{-3mm} & \hspace{-2mm} {\hat{I_p}}_y\\ -{\hat{I_p}}_y \hspace{-3mm} & \hspace{-2mm} {\hat{I_p}}_x \end{pmatrix} \hspace{-0.75mm}
 \hspace{-1mm}
\begin{pmatrix} \Delta{V_p}_x \\ \Delta{V_p}_y \end{pmatrix} \hspace{-1.5mm}
\end{array}
\hspace{-3mm}
\end{equation}

The perturbed voltage constraints are provided in \eqref{VolCon} for bus $b$. The perturbed current constraint of line $l$ is presented in \eqref{CurCon}. The sending and receiving ends are given by subscripts 1 and 2, respectively.
Tap positions and \textsc{cb} steps should be set between the minimum and maximum values.
For \textsc{svr}s and \textsc{rr}s, capacity limits are given in \eqref{SVRLimit} and \eqref{RRLimit}, respectively.
For photovoltaic units, \eqref{RRLimit2} gives the maximum power angle constraint to avoid high harmonic distortions. For a doubly-fed induction wind generator, the reactive power cannot be lower than a specified value \eqref{RRLimit3}.
For dispatchable \textsc{der}s, \eqref{DERLimit} gives the perturbed capacity constraints.
\begin{equation}
\label{VolCon}
\begin{array}{l}
\displaystyle

\left(V_b^{min}\right)^2 
-
\left|\hat{V}_b\right|^2
\leq
2{\hat{V_b}}_x\Delta{V_b}_x
+
2{\hat{V_b}}_y\Delta{V_b}_y
\\
\displaystyle

2{\hat{V_b}}_x\Delta{V_b}_x
+
2{\hat{V_b}}_y\Delta{V_b}_y
\leq
\left(V_b^{max}\right)^2
-
\left|\hat{V}_b\right|^2

\end{array}
\hspace{-3mm}
\end{equation}
\begin{equation}
\label{CurCon}
\begin{array}{l}
\displaystyle
\hspace{-2mm}
2\left({Y_l}_x{\hat{I_l}}_x 
+
{Y_l}_y{\hat{I_l}}_y\right) 

\left(\Delta {V_1}_x-\Delta {V_2}_x\right) +

\\
\displaystyle
\hspace{-2mm}
2\left({Y_l}_x{\hat{I_l}}_y
-
{Y_l}_y{\hat{I_l}}_x\right)

\left(\Delta {V_1}_y-\Delta {V_2}_y\right)

\leq 
\left(I_l^{max}\right)^2
-
\left|\hat{I}_l\right|^2
\end{array}
\hspace{-2mm}
\end{equation}
\begin{equation}
\label{SVRLimit}
\displaystyle
-S_{svr}^{n} - \hat{Q}_g^{svr} \leq \Delta Q_g^{svr}\leq S_{svr}^{n} - \hat{Q}_g^{svr}
\end{equation}
\begin{equation}
\label{RRLimit}
\displaystyle
\pm \hat{Q}_g^{rr} \pm \Delta Q_g^{rr}
\leq
\sqrt{\left(S_{rr}^{n}\right)^2
-
\left(\hat{P}_g^{rr}\right)^2}
\end{equation}
\begin{equation}
\label{RRLimit2}
\displaystyle
-tan(\alpha^{\text{max},pv}) \hat{P}_g^{pv}
\leq
\hat{Q}_g^{pv}+
\Delta Q_g^{pv}
\leq
tan(\alpha^{\text{max},pv}) \hat{P}_g^{pv}
\end{equation}
\begin{equation}
\label{RRLimit3}
\displaystyle
Q^{\text{min},wind}
\leq
\hat{Q}_g^{wind}+
\Delta Q_g^{wind} \hspace{-1.5mm}
\end{equation}
\begin{equation}
\label{DERLimit}
\displaystyle
\hspace{1mm}
2 \hat{P}_g^{der} \hspace{-1mm}
\Delta P_g^{der} \hspace{-1mm}
+ \hspace{-1mm}
2 \hat{Q}_g^{der} \hspace{-1mm}
\Delta Q_g^{der} \hspace{-1mm}
\leq \hspace{-1mm}
\left(S_{der}^{n}\right)^2 \hspace{-1mm}
- \hspace{-1mm}
\left(\hat{P}_g^{der}\right)^2 \hspace{-1mm}
- \hspace{-1mm}
\left(\hat{Q}_g^{der}\right)^2 \hspace{-2mm} 
\end{equation}

\textsc{tra} was first proposed to solve unconstrained nonlinear problems, then to solve the problems with equality constraints and finally to handle the simple bounds on optimization variables \cite{Songqiang2018}. The inequality constraints are first converted to equality constraints using slack variables. Constraint \eqref{VolCon} is selected for demonstration. The equality constraints and simple bound on slack variables $\epsilon^+$ and $\epsilon^-$ are given in \eqref{EQSBound}.
\begin{equation}
\label{EQSBound}
\begin{array}{l}
\displaystyle

2{\hat{V_b}}_x\Delta{V_b}_x 
+
2{\hat{V_b}}_y\Delta{V_b}_y +\epsilon^+_b=
\left(V_b^{max}\right)^2
-|\hat{V}_b|^2

\\
\displaystyle
2{\hat{V_b}}_x\Delta{V_b}_x
+
2{\hat{V_b}}_y\Delta{V_b}_y-\epsilon^-_b=
\left(V_b^{min}\right)^2
-|\hat{V}_b|^2

\\
\displaystyle
0 \leq \epsilon^+_b, \epsilon^-_b

\end{array}
\hspace{-3mm}
\end{equation}
\section{Controllable Devices and Network Models}\label{SecII}
\subsection{Transformers and Upstream Network Models}\label{SecA}
The \textsc{oltc} control affects the transformer model.
It is assumed that the tap changer has been installed on the primary winding, i.e., $N_s$ is constant. The transformer core is assumed to remain unsaturated. As the tap position ($tap$) increases, $N_p$ and the turn ratio ($r$) increase. In the nominal tap position ($tap^n$=0) the turn ratio (in pu.) is 1 ($r^n$=1).
Tap changing operations change $Z_p$ proportional to $N_p$. $Z_s$, $X_M$ and $R_c$ viewed from the secondary terminal do not change. 

\begin{figure}[!t]
  \centering
  \includegraphics[width=0.9\columnwidth,clip=true]{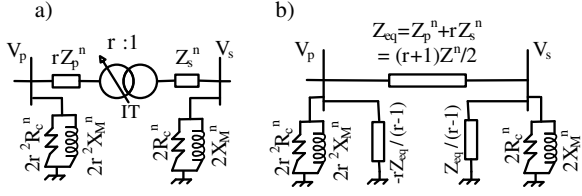}
  \caption{a) Transformer model under \textsc{oltc}, b) equivalent pi circuit.}\label{Trans1}
\end{figure}

\begin{figure}[!t]
  \centering
  \includegraphics[width=0.60\columnwidth,clip=true]{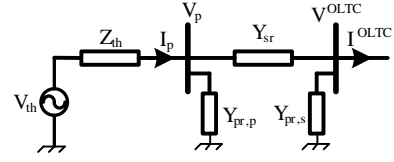}
  \caption{Upstream network and transformers' models.}\label{Upstream}
\end{figure}

Fig. \ref{Trans1} (a), shows the equivalent circuit of a transformer, where IT is an ideal transformer. Fig. \ref{Trans1} (b) shows the equivalent per unit $pi$ circuit. It is assumed that for a well manufactured transformer $Z_{p,t}^{n}=Z_{s,t}^{n}=Z_t^n$/2 \cite{Pouladi2019}.
\textcolor[rgb]{0,0,0}{For $N_t$ parallel transformers indexed by $t$, each impedance in the resultant equivalent $pi$ model is found by aggregating the respective impedances in the $pi$ model of all transformers, i.e, 1/$Z_{sr}$=$\sum^{N_t}_{t=1}1/{Z_{sr}}_t$, 1/$Z_{pr,p}$=$\sum^{N_t}_{t=1}1/{Z_{pr,s}}_t$ and 1/$Z_{pr,s}$=$\sum^{N_t}_{t=1}1/{Z_{pr,p}}_t$ (see Fig. \ref{Upstream}).} This model can be readily used in power flow studies.

The upstream system is modeled by the equivalent Thevenin model. However, it is hard to accurately estimate $V_{th}$ and  $Z_{th}$.
By tracking the \textcolor[rgb]{0,0,0}{variations} of measured (or estimated) $V_p$ and $I_p$, $V_{th}$ and  $Z_{th}$ can be found, if these variations are caused dominantly by a change in the downstream network. However, during the normal operation, \textcolor[rgb]{0,0,0}{$V_p$ and $I_p$ }change gradually
and it is very hard to understand if the source of these changes is in the upstream or downstream systems.

Fig. \ref{Upstream} shows the model used for the upstream system and parallel transformers.
It was proposed in \cite{Bahadornejad2014}, to deliberately change the control variables in downstream network to find $Z_{th}$.
Here, the Thevenin model is found by comparing the measured values of $V_p$ and $I_p$, before and after applying some of the changes proposed by the scheduling algorithm.
The changes should be significant enough to cancel the effects of measurement errors.
The measurement instants should be as close as possible.
The Thevenin model can be found using \eqref{Thevinin_model_b}. \textcolor[rgb]{0,0,0}{Superscripts $\xi$ takes the values 0, 1, and 2 to indicate the values before applying the changes proposed by the scheduling algorithm, after applying the first change and after applying the second change, respectively.}
In \eqref{Thevinin_model_b}, the measured values are distinguished using a bar upon them. \textcolor[rgb]{0,0,0}{There are six variables, i.e., $V_{th}$, $Z_{th}$, $\delta_{Z_{th}}$, $\delta_{V_{p}^0}$, $\delta_{V_{p}^1}$ and $\delta_{V_{p}^2}$ and three equations of type \eqref{Thevinin_model_b} which are rewritten in six equations separating the real and imaginary parts. Therefore, $V_{th}$ and $Z_{th}$ can be found.}
\begin{equation}
\label{Thevinin_model_b}
\displaystyle
\left|V_{th}\right|^{\angle{0}}=\overline{\left|V_{p}^\xi \right|}^{\angle{\delta_{V_{p}^\xi}}}   +Z_{th}^{\angle{\delta_{Z_{th}}}}.\overline{\left|I_{p}^\xi \right|}^{\angle{\delta_{V_{p}^\xi}}+\overline{\varphi_{I_{p}^\xi,V_{p}^\xi}}}
\end{equation}

The algorithm presented in steps 1-4, shows how to update the upstream system model for $\nu^{th}$ scheduling period. As will be seen, the proposed scheduling method is fast enough to update the scheduling results accordingly. Tolerances $\epsilon_Z$ and $\epsilon_V$ are set to achieve an acceptable solutions accuracy.
\begin{enumerate}
	\item Let $Z_{th}^{\nu}=Z_{th}^{\nu-1}$, find $\left|V_{th}^{\nu}\right|$ using the measured $V_p$ and $I_p$ and \eqref{Thevinin_model_b} ($\left|V_p\right|$, $\left|I_p\right|$, $\varphi_{I_{p},V_{p}}$, $Z_{th}^{\nu}$ and $\delta_{Z_{th}^{\nu}}$ are known).
	\item Run the scheduling algorithm, apply two cheapest changes and measure $V_p$ and $I_p$ after each change.  
	\item Using $V_p$ and $I_p$ measured in these three instants and \eqref{Thevinin_model_b} find  $\left|V_{th}^{\nu,new}\right|$ and $Z_{th}^{\nu,new}$.
	\item If $\left| \left|V_{th}^{\nu,new} \right| - \left|V_{th}^{\nu}\right| \right| \geq \epsilon_V$
or $\left| Z_{th}^{\nu,new} - Z_{th}^{\nu} \right| \geq \epsilon_Z$, run the scheduling algorithm again and apply all the changes.
\end{enumerate}

The relationship between the voltage and current at the secondary bus (\textsc{oltc} bus) and the \textsc{oltc} control variables ($tap$) is shown in \eqref{OLTC_Control}. $C(r)$ and $D(r)$ are found according to Fig. \ref{Upstream}. Equation \eqref{r_tap} shows the relationship between $r_t$ and $tap_t$. In the perturbed relationship between $I^{\text{OLTC}}$, $V^{\text{OLTC}}$ and vector $tap$ \eqref{Linear_OLTC}, matrices $A$ and $B$ are defined in \eqref{Linear_OLTC_A} and \eqref{Linear_OLTC_B}, respectively.
Superscript \textsc{oltc} has been removed for brevity.
\begin{equation}
\label{OLTC_Control}
\displaystyle
I^{\text{OLTC}}=C(r)V^{\text{OLTC}}+D(r)V_{th}
\end{equation}
\begin{equation}
\label{r_tap}
\displaystyle
r_t=1+tap_t{\Delta}U_t
\end{equation}
\begin{equation}
\label{Linear_OLTC}
\displaystyle
\begin{pmatrix} \Delta{I_x^{\text{OLTC}}} \\ \Delta{I_y^{\text{OLTC}}} \end{pmatrix}=
A^{\text{OLTC}}_{2\times{2}}\begin{pmatrix} \Delta{V_x^{\text{OLTC}}} \\ \Delta{V_y^{\text{OLTC}}} \end{pmatrix}+B^{\text{OLTC}}_{2\times{N_t}}\Delta{tap}
\end{equation}
\begin{equation}
\label{Linear_OLTC_A}
\displaystyle
A^\text{OLTC}=
\begin{pmatrix} C_x & -C_y\\ C_y & C_x \end{pmatrix}
\end{equation}
\begin{equation}
\label{Linear_OLTC_B}
\begin{array}{l}
\displaystyle

B^\text{OLTC} 
= 
\begin{pmatrix}
\hat{V}_x\frac{\partial{C_x}}{\partial{tap}}
+ 
{V_{th}}\frac{\partial{D_x}}{\partial{tap}}
-
\hat{V}_y\frac{\partial{C_y}}{\partial{tap}}
\\
\hat{V}_y\frac{\partial{C_x}}{\partial{tap}}
+
\hat{V}_x\frac{\partial{C_y}}{\partial{yap}}
+
{V_{th}}\frac{\partial{D_y}}{\partial{tap}}
\end{pmatrix}
_{2\times{N_t}}
\end{array}
\end{equation}
\subsection{SVRs, DERs, RRs and CBs}\label{SecB}
The perturbed model is presented here for continuous control devices, e.g., \textsc{svr}s, \textsc{rr}s and \textsc{der}s, in power control (\textsc{pq}) mode. The voltage control mode is discussed in \ref{SecE}. Considering $S=VI^*$, the perturbed model of each continuous control device is presented in \eqref{Contin_Linear}. $I$ is the current injected by this control device. For the devices which cannot control their active power, $\Delta P_g$=0. It means the regarding line should be eliminated from $A^c$.
The reactive power that \textsc{cb}s inject to the network is a function of their impedances and their voltages.
The perturbed model of each \textsc{cb} is given in \eqref{CBs_Linear}.
\begin{equation}
\label{Contin_Linear}
\begin{array}{l}
\displaystyle
\begin{pmatrix} \Delta{I_x} \\ \Delta{I_y}\end{pmatrix}
=
\overbrace{-\begin{pmatrix} \hat{V}_x & \hat{V}_y\\ \hat{V}_y & -\hat{V}_x \end{pmatrix}^{-1} \begin{pmatrix} \hat{I}_x & \hat{I}_y\\ -\hat{I}_y & \hat{I}_x \end{pmatrix}}^{A^{c}_{2\times{2}}}

\begin{pmatrix} \Delta{V_x} \\ \Delta{V_y} \end{pmatrix}
\vspace{2mm}
\\
\displaystyle
\hspace{12mm} +
\underbrace{\begin{pmatrix} \hat{V}_x & \hat{V}_y\\ \hat{V}_y & -\hat{V}_x \end{pmatrix}^{-1}}_{B^{c}_{2\times{2}}}
\begin{pmatrix} \Delta{P}_g \\ \Delta{Q}_g \end{pmatrix}
\end{array}
\end{equation}
\begin{equation}
\label{CBs_Linear}
\begin{array}{l}
\displaystyle
\hspace{-3mm}
\begin{pmatrix} \Delta{I_x} \\ \Delta{I_y}\end{pmatrix} \hspace{-1mm}
= \hspace{-1mm}
\underbrace{\begin{pmatrix} 0 & y^{st}\hat{st} \\ -y^{st}\hat{st} & 0 \end{pmatrix}}_{A^{cb}_{2\times 2}}
\begin{pmatrix} \Delta{V_x} \\ \Delta{V_y} \end{pmatrix} \hspace{-1mm}
+\hspace{-1mm}
\underbrace{\begin{pmatrix} y^{st}\hat{V}_y \\ -y^{st}\hat{V}_x \end{pmatrix}}_{B^{cb}_{2\times 1}}
\Delta{st}
\end{array}
\hspace{-1mm}
\end{equation}
\vspace{-2mm}
\subsection{Load Model}\label{SecD}
In some studies the voltage profile improvement was considered as one of the objectives.
In contrast, the voltage-dependent nature of loads can be deemed as an opportunity to reduce the cost by managing the voltage levels.
A quadratic function can be used to approximate the \textcolor[rgb]{0,0,0}{steady-state} relationship between the load levels and bus voltages \cite{Nouri2017}.
\textcolor[rgb]{0,0,0}{Such model is referred to as ZIP load model as it combines the constant impedance (Z component), constant current (I component) and constant power (P component) characteristics of the loads \eqref{ZIP}.}

\textcolor[rgb]{0,0,0}{The proposed method can accommodate any load model with desired level of accuracy and complexity. However, for the sake of the simplicity of presentation, the ZIP load model is replaced with a ZP model. Within the typical range of voltages in the steady state conditions, the accuracies of ZIP and ZP models are quite close and the ZIP model can be reduced to a ZP model \eqref{ZIP2ZP}.
Replacing \eqref{ZIP2ZP} in \eqref{ZIP} and comparing the resultant equations to \eqref{ZP}, $\zeta'_p$=$\zeta_p$+$\mu_p$/2 and $\kappa'_p$=$\kappa_p$+$\mu_p$/2.}

\textcolor[rgb]{0,0,0}{Coefficients of this quadratic model are not fixed, since the combination of load components is varying from time to time.
A fixed load model cannot be applied to optimize the system cost (by reducing the demand and power loss based on the voltage-dependent nature of the loads). The parameters of the load model should be kept up to date. 
Here, an adaptive perturbed load model is presented.
Without an accurate load model, the expected energy saving cannot be realized.
Parameters of the ZP model are $\zeta'_p$, $\kappa'_p$, $\zeta'_q$, $\kappa'_q$, $P_{d_0}$ and $Q_{d_0}$.
For $\left|V\right|$=$V_0$, $P_d$=$P_{d_0}$ and $Q_d$=$Q_{d_0}$. Therefore, $\zeta'_p$+$\kappa'_p$=$\zeta'_q$+$\kappa'_q$=1 and independent parameters include $\zeta'_p$, $\zeta'_q$, $P_{d_0}$ and $Q_{d_0}$.
To update the load model, it is sufficient to update these independent parameters.}

\textcolor[rgb]{0,0,0}{With just the measured or estimated $P_d$, $Q_d$ and $V$ at each bus before applying the scheduling technique, there are only two equations (the ones presented in \eqref{ZP}) to extract these four independent parameters.
Therefore, another set of measurements for $P_d$, $Q_d$ and $V$ is required.
In the first step, the scheduling framework uses the latest updated load parameters and solves the \textsc{dsp}. 
The second set of $P_d$, $Q_d$ and $V$ is measured after applying the changes proposed by the scheduling framework.
The method is similar to the one used for updating upstream model in \ref{SecA}.
If the accurate load model parameters are significantly different from those used by the scheduling framework, the scheduling problem is solved again.
Using \eqref{LoadAprPwr} and \eqref{ZP}, the same perturbed equation as \eqref{Contin_Linear} is found for the loads. $A^{load}$ is given in \eqref{Load_Linear_2} and $B^{load}$=0.}
\begin{equation}
\label{ZIP}
\begin{array}{l}

\displaystyle
\frac{P_d}{P_{d_0}}=\zeta_p\left(\frac{|V|}{V_0}\right)^2+\mu_p\left(\frac{|V|}{V_0}\right)+\kappa_{p} \\

\displaystyle
\frac{Q_d}{Q_{d_0}}=\zeta_q\left(\frac{|V|}{V_0}\right)^2+\mu_q\left(\frac{|V|}{V_0}\right)+\kappa_q

\end{array}
\end{equation}
\begin{equation}
\label{ZIP2ZP}
\begin{array}{l}

\displaystyle
\textcolor[rgb]{0,0,0}{\frac{\left|V\right|}{V_0}  \approx 0.5 \left(1+\frac{\left|V\right|^2}{V_0^2}\right)}

\end{array}
\end{equation}
\begin{equation}
\label{ZP}
\begin{array}{l}

\displaystyle
\frac{P_d}{P_{d_0}}=\zeta'_p \frac{|V|^2}{V_0^2}+\kappa'_p, 

\hspace{7mm}

\frac{Q_d}{Q_{d_0}}=\zeta'_q\frac{|V|^2}{V_0^2}+\kappa'_q

\end{array}
\end{equation}
\begin{equation}
\label{LoadAprPwr}
\begin{array}{l}
\displaystyle

\textcolor[rgb]{0,0,0}{\hat{V}_x \Delta{I}_x+\hat{V}_y \Delta{I}_y+\hat{I}_x \Delta{V}_x+\hat{I}_y \Delta{V}_y=-\Delta P_d} \\
\displaystyle
\textcolor[rgb]{0,0,0}{\hat{V}_y \Delta{I}_x-\hat{V}_x \Delta{I}_y-\hat{I}_y \Delta{V}_x+\hat{I}_x \Delta{V}_y=-\Delta Q_d}

\end{array}
\end{equation}
%
\begin{equation}
\label{Load_Linear_2}
\begin{array}{l}
\displaystyle
\hspace{-3.5mm}
A^{load}
=
\begin{pmatrix} \hat{V}_x \hspace{-2mm} & \hspace{-1.5mm} \hat{V}_y \\ \hat{V}_y \hspace{-1.5mm} & \hspace{-2mm} -\hat{V}_x \end{pmatrix} ^{\hspace{-1.5mm} -1} \hspace{-2mm}
\vspace{1mm}

\begin{pmatrix}
\hspace{-0.5mm}
\frac{2\zeta'_p P_{d_0} \hspace{-0.5mm} \hat{V}_x}{V_0^2} \hspace{-1mm} - \hspace{-1mm} \hat{I}_x \hspace{-1.5mm}
& \hspace{-1.5mm} \frac{2\zeta'_p P_{d_0} \hspace{-0.5mm} \hat{V}_y}{V_0^2} \hspace{-1mm} - \hspace{-1mm} \hat{I}_y \hspace{-1.5mm}  \\

\hspace{-0.5mm}
\frac{2\zeta'_q Q_{d_0} \hspace{-0.5mm} \hat{V}_x}{V_0^2} \hspace{-1mm} + \hspace{-1mm} \hat{I}_y \hspace{-2mm}
& \hspace{-1mm} \frac{2\zeta'_q Q_{d_0} \hspace{-0.5mm} \hat{V}_y}{V_0^2} \hspace{-1mm} - \hspace{-1mm} \hat{I}_x \hspace{-1.5mm}
\end{pmatrix}
\end{array}
\end{equation}
\subsection{Network Model}\label{SecIII}
\begin{figure}[!t]
  \centering
  \includegraphics[width=0.8\columnwidth,clip=true]{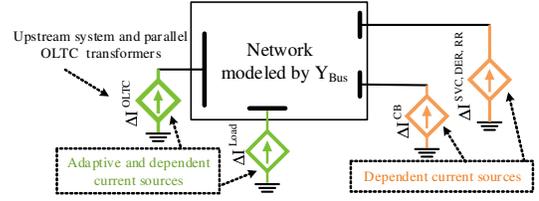}
	\caption{Dependent current source representation of the perturbed models.}\label{CSR}
\end{figure}

The models developed so far are combined with the network model to form the final perturbed formulation.
The system network is modeled using bus admittance matrix and controllable devices are modeled as a \emph{controllable dependent current sources} (Fig. \ref{CSR}), i.e., the current is controlled by changing the control variables and also depends on the bus voltage.

The network model is presented in \eqref{NetEq}. The complete perturbed model is presented in \eqref{LinModel0}, in which the perturbed currents has been replaced in \eqref{NetEq} using the right hand side of the equations developed for controllable devices and loads. The perturbed model is rearranged in \eqref{LinModel}. The entries of matrices $A$ and $B$ are found by aggregating the perturbed models developed for the controllable devices and loads.
\begin{equation}
\label{NetEq}
\begin{array}{l}
\displaystyle

\begin{pmatrix} \Delta{I}_x \\ \Delta{I}_y \end{pmatrix}
=
\begin{pmatrix} {Y_\text{Bus}}_x & -{Y_\text{Bus}}_y
\\
{Y_\text{Bus}}_y & {Y_\text{Bus}}_x \end{pmatrix}_{2N_\text{B}\times{2N_\text{B}}}

\begin{pmatrix} \Delta{V}_x \\ \Delta{V}_y \end{pmatrix}
\end{array}
\end{equation}
\begin{equation}
\label{LinModel0}
\begin{array}{l}
\displaystyle
A
\begin{pmatrix} \Delta{V}_x \\ \Delta{V}_y \end{pmatrix}
+
\left[ B\right]_{2N_\text{B}\times{N_\text{W}}}\left[\Delta{W} \right] 
=
Y
\begin{pmatrix} \Delta{V}_x \\ \Delta{V}_y \end{pmatrix}
\end{array}
\end{equation}
\begin{equation}
\label{LinModel}
\begin{array}{l}
\displaystyle
\begin{pmatrix} \Delta{V}_x \\ \Delta{V}_y \end{pmatrix}
=
\left(Y_\text{Bus}-A\right)^{-1}\left[B\right]\Delta{W} 
\end{array}
\end{equation}
\subsection{Voltage Control Mode}\label{SecE}
For voltage control mode (denoted by superscript $vc$), the perturbed model can be found using \eqref{VCM0}.
This equation is used along with the active power part of $S=VI^*$ to build the linearized model of \eqref{VCM}.
The reactive part of $S=VI^*$ is used to develop the capacity constraint. For the devices that cannot change their active power $\Delta{P}_g$ is replaced by 0.
\begin{equation}
\label{VCM0}
\begin{array}{l}
\displaystyle
\hspace{-3mm}
(\hat{V}_x^{vc}+\Delta{V}_x^{vc})^2 \hspace{-1mm}
+ \hspace{-1mm}
(\hat{V}_y^{vc}+\Delta{V}_y^{vc})^2 \hspace{-1mm}
= \hspace{-1mm}
(\left|\hat{V}^{vc}\right|+\Delta{\left|V^{vc}\right|})^2
\end{array}
\end{equation}
\begin{equation}
\label{VCM}
\begin{array}{l}
\hspace{-4.75mm}
\displaystyle
\begin{pmatrix} \hat{V}_x^{vc} \hspace{-2.75mm} & \hspace{-2.25mm} \hat{V}_y^{vc} \hspace{-0.5mm} \\ \hspace{-1.5mm} 0 \hspace{-2.75mm} & \hspace{-2.25mm} 0 \hspace{-2mm} \end{pmatrix}\hspace{-1.75mm}
\begin{pmatrix} \Delta{I}_x^{vc} \\ \Delta{I}_y^{vc} \end{pmatrix}\hspace{-1.25mm}
+\hspace{-1.25mm}
\begin{pmatrix} \hat{I}_x^{vc} \hspace{-2mm} & \hspace{-2mm} \hat{I}_y^{vc} \\ \hat{V}_x^{vc} \hspace{-2mm} & \hspace{-2mm} \hat{V}_y^{vc} \end{pmatrix}\hspace{-1.75mm}
\begin{pmatrix} \Delta{V}_x^{vc} \\ \Delta{V}_y^{vc} \end{pmatrix} \hspace{-1.5mm}
=\hspace{-1.5mm}
\begin{pmatrix} \Delta{P}_g^{vc} \\ \Delta{\left|V^{vc}\right|} \end{pmatrix} \hspace{-3.5mm}
\end{array}
\end{equation}

The matrix of the coefficients of perturbed current vector in \eqref{VCM} is not invertible.
To solve the issue, the effects of these control devices are incorporated in \eqref{LinModel0} using vector $\Delta I^{vc}$. The resultant equation is given in \eqref{LinMode_and_VCM}. The perturbed currents in \eqref{VCM} are replaced with linearized expressions in terms of the perturbed voltages and control variables using \eqref{LinMode_and_VCM}. After rearranging the resultant equation, two fresh equations are found for the perturbed voltages in terms of $[\Delta{W}]$ including ($\Delta{\left|V^{vc}\right|}$ and $\Delta{P}_g^{vc}$). The equations of \eqref{LinMode_and_VCM} that contain $\Delta{I}_x^{vc}$ and $\Delta{I}_y^{vc}$ are replaced with this two fresh equations.
\begin{equation}
\label{LinMode_and_VCM}
\begin{array}{l}
\displaystyle
A.\begin{pmatrix} \Delta{V}_x \\ \Delta{V}_y \end{pmatrix}
+
B.\Delta{W} 
+
\begin{pmatrix} 0 \vspace{-2mm} \\ \vspace{-1mm} \vdots \\ \vspace{-2mm} \Delta{I}_x^{vc} \\ \vspace{-1mm} \vdots \\ \Delta{I}_y^{vc} \vspace{-2mm}
\\ \vspace{-2mm} \vdots \\ \vspace{-3mm} \\ \vspace{-1mm} 0 \end{pmatrix}
=
Y
.
\begin{pmatrix} \Delta{V}_x \\ \Delta{V}_y \end{pmatrix}
\end{array}
\end{equation}
\section{How to Expedite the Solution}\label{SecIIIC}
To further expedite the proposed \textsc{minlp} methodology, at each node of \textsc{bc} algorithm, a simplified problem is first solved. This may obviate the need of applying \textsc{tra} to solve the regarding integer relaxed problem. The objective function and constraints of this simplified problem are the linearized cost function and constraints around $\hat{\Psi}$.
Constraints, \eqref{VolCon}, \eqref{CurCon} and \eqref{DERLimit} are originally nonlinear.
In \eqref{VolCon}, the negligible term $(\Delta V_x)^2+(\Delta V_y)^2$ has been omitted. This does not affect the solution optimality. In \eqref{CurCon} and \eqref{DERLimit}, the always positive terms $\left(\Delta I_x\right)^2+\left(\Delta I_y\right)^2$ and $\left(\Delta P_g\right)^2+\left(\Delta Q_g\right)^2$ are neglected. This relaxes these constraints to some extent.

According to subsection \ref{IE}, the accurate objective function is concave up, i.e., the linearized objective function is always lower than the accurate objective function, some of the linearized constraints are weaker than the regarding non-linear constraints and the other constraints are originally linear.
Therefore, if at a certain node, the solution gained by \textsc{lp} is dominated by the best \textsc{minlp} solution seen so far or is infeasible, the accurate nonlinear programming solution is also dominated or infeasible and the current branch is not to be continued further. Under this setup, \textsc{lp} helps to find the solution of the intermediate nodes and to faster make the decision at the end nodes without compromising the solution optimality.
For the candidate solutions, i.e., the solutions for which the value of the linearized objective function is less than ``$z^*$'', \textsc{tra} is used to further inspect the solution feasibility and optimality. $z^*$ is the value of the objective function for the best \textsc{minlp} solution already found.
For the nodes at which the application of \textsc{tra} is inevitable, this \textsc{lp} is not a redundant step and its solution is used as the starting point of \textsc{tra}.

\begin{figure}[!t]
  \centering
  \includegraphics[width=0.80\columnwidth, clip=true]{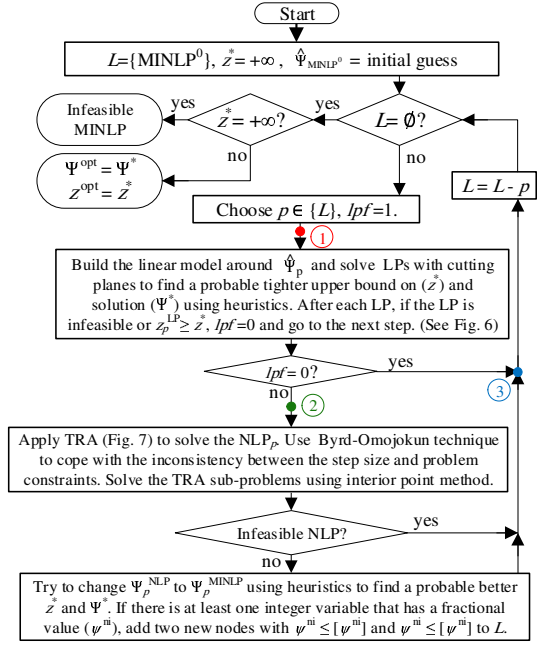}
  \caption{\textcolor[rgb]{0,0,0}{Comprehensive flowchart of the proposed solution methodology.}}\label{Flow1}
\end{figure}

\begin{figure}[!t]
  \centering
  \includegraphics[width=0.75\columnwidth, clip=true]{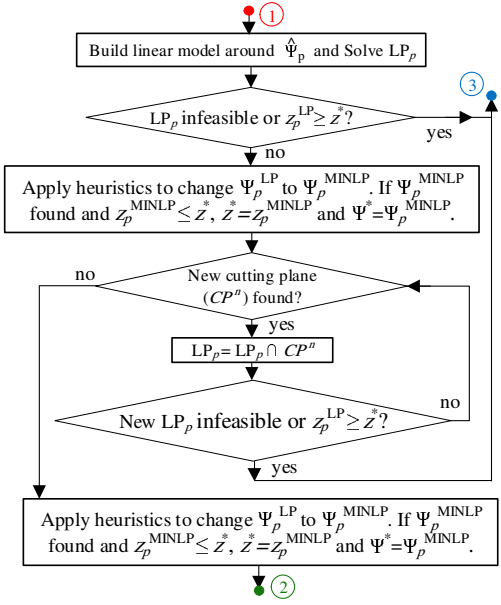}
  \caption{\textcolor[rgb]{0,0,0}{Proposed expediting technique.}}\label{Flow2}
\end{figure}

\begin{figure}[!t]
  \centering
  \includegraphics[width=0.85\columnwidth, clip=true]{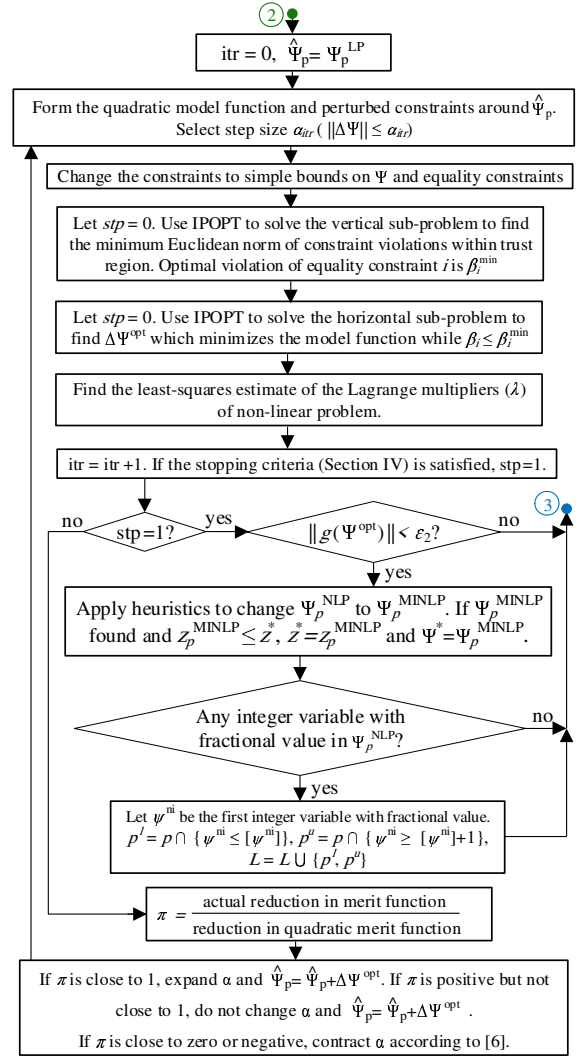}
  \caption{\textcolor[rgb]{0,0,0}{Steps of \textsc{tra} for solving \textsc{nlp} problems.}}\label{Flow3}
\end{figure}

\textcolor[rgb]{0,0,0}{The comprehensive flowchart of Fig. \ref{Flow1} presents an overview of the proposed globally convergent \textsc{minlp} solution methodology.
The proposed expediting technique, i.e., the steps between point \circled{1} and points \circled{2} and \circled{3} in Fig. \ref{Flow1},  is further outlined in Fig. \ref{Flow2} and will be discussed later.
In Fig. \ref{Flow1} the steps of applying \textsc{tra} to solve the \textsc{nlp} problems during the branching process of \textsc{bc} are presented between points \circled{2} and \circled{3}.
These steps are also outlined in Fig. \ref{Flow3}.
$L$ (which is indexed by $p$) is the set of all \textsc{minlp} problems that should be solved during the branching process of \textsc{bc} technique, $\text{MINLP}^0$ is the original problem and $\Psi^*$ is the best integer feasible solution found so far. In these flowcharts, $z_p$ is the value of objective function after solving problem $p$. For the optimal solution point, $z=z^\text{opt}$ and $\Psi=\Psi^\text{opt}$. The optimal solution of \textsc{lp} problem $p$, optimal solution of \textsc{nlp} problem $p$ and a feasible solution for \textsc{minlp} problem $p$ are given by $\Psi_p^\text{LP}$, $\Psi_p^\text{NLP}$ and $\Psi_p^\text{MINLP}$, respectively.}

\textcolor[rgb]{0,0,0}{During the solution process a heuristic approach is applied to change the solutions of the \textsc{lp} and \textsc{nlp} problems to feasible \textsc{minlp} solutions when possible.
In this approach, the values of integer variables are first rounded-off to the nearest integer values according to \cite{Dey2018}.
\textsc{tra} is next applied to find the values of continuous variables.
Using this approach, sometimes a tighter upper bound is found for the objective function.
This step is not outlined in the flowchart and is referred to as ``changing the $\Psi_p^\text{LP}$/$\Psi_p^\text{NLP}$ to a $\Psi_p^\text{MINLP}$ using heuristics''.
If this step succeeds to find a $\Psi_p^\text{MINLP}$ and $z^\text{MINLP}$ is lower than $z^*$, $z^*$ is replaced with $z^\text{MINLP}$ as a tighter upper bound.}

\textcolor[rgb]{0,0,0}{For the simplified problem (Fig. \ref{Flow2}) at a certain node during the branching process, the proposed method first drops the integerality constraints and solves the associated \textsc{lp} problem.
The solution will be a vertex of the convex polytope consisting of all feasible solutions.
If this vertex is not an integer feasible solution, the method first tries to change this solution to a \textsc{minlp} feasible solution using the proposed heuristic approach.}

\textcolor[rgb]{0,0,0}{The proposed algorithm also uses cutting planes to expedite the solution of simplified problem. 
After finding the optimal vertex using \textsc{lp}, the method finds a hyperplane ($cp^n$ in Fig. \ref{Flow2}) with this vertex on one side and all integer feasible solution on the other side.
$cn^n$ is then added as a new linear constraint to exclude this integer infeasible vertex.
This new \textsc{lp} is solved and the process is repeated until an integer solution is found or no more cutting planes can be found (Fig. \ref{Flow2}).
More explanation on developing these cutting planes was provided in \cite{Dey2018}.}

\textcolor[rgb]{0,0,0}{The implementation of \textsc{tra} to solve the \textsc{nlp}s during the branching process is outlined in Fig. \ref{Flow3}.
The steps were also discussed in Section \ref{sec_1}. After changing the constraints of the \textsc{nlp} problem to the simple bounds on the optimization variables ($\Psi$) and equality constraints (see subsection \ref{TRA0}), the problem can be formulated as \eqref{nlp}. The quadratic model function and perturbed constraints are built for each \textsc{tra} sub-problem based on subsection \ref{TRA0}.
In Fig. \ref{Flow3}, $\left\|W\right\|$ gives the Euclidean norm of vector $W$ and $[\psi]$ is the integer part of real variable $\psi$.
Each sub-problem is divided into a vertical sub-problem and a horizontal sub-problem with the functionalities presented in Fig. \ref{Flow3}.
In the vertical sub-problem, the objective is to minimize the Euclidean norm of constraint violations within the trust region ($||\Delta \Psi|| \leq \alpha$). The result of this step includes the optimal constraint violation $\beta_i^\text{min}$ for each constraint $g_i(\Psi)=0$.
Stopping criteria include the Lagrange optimality and constraints' satisfaction conditions, which are provided in \eqref{StopCr1} and \eqref{StopCr2}, respectively.
After each \textsc{tra} sub-problem, if \eqref{StopCr1} and \eqref{StopCr2} are simultaneously satisfied or $itr \geq itr^{max}$, \textsc{tra} is stopped. The number of \textsc{tra} sub-problems solved in order to solve this \textsc{nlp} is given by $itr$. In \eqref{StopCr1}, vector $\lambda$ gives the Lagrange multipliers of the equality constraints in \eqref{nlp}.
These multipliers are not computed by \textsc{tra}. They are found using a least-squares estimate based on \cite{BAHRAMI2020}. Finally, to decide on the step size for the next iteration, parameter $\pi$ is used according to Fig. \ref{Flow3} and \cite{BAHRAMI2020}. The merit function $\phi(\Psi)$ is provided in \eqref{mrt}. $\eta \geq 1$ is a penalty parameter that weights constraint satisfaction against objective minimization.}
\begin{equation}
\label{nlp}
\begin{array}{l}
\displaystyle
\textcolor[rgb]{0,0,0}{\text{min} \hspace{4mm} f(\Psi)}\\
\displaystyle
\textcolor[rgb]{0,0,0}{\text{s.t.:} \hspace{5mm} g(\Psi)=0}\\
\displaystyle
\textcolor[rgb]{0,0,0}{\hspace{10mm} \Psi^{min} \leq \Psi \leq \Psi^{max}}
\end{array}
\end{equation}
\begin{equation}
\label{StopCr1}
\begin{array}{l}
\displaystyle
\textcolor[rgb]{0,0,0}{\left\| \nabla f(\Psi)+ \nabla g(\Psi)^T \lambda \right\|<\epsilon_1}
\end{array}
\end{equation}
\begin{equation}
\label{StopCr2}
\begin{array}{l}
\displaystyle
\textcolor[rgb]{0,0,0}{\left\| g(\Psi) \right\|<\epsilon_2}
\end{array}
\end{equation}
\begin{equation}
\label{mrt}
\begin{array}{l}
\displaystyle
\textcolor[rgb]{0,0,0}{\phi(\Psi)=f(\Psi)+ \eta \left\| g(\Psi) \right\|}
\end{array}
\end{equation}
\subsection{Illustrative Example}\label{IE}
\begin{figure}[!t]
  \centering
  \includegraphics[width=0.65\columnwidth,clip=true]{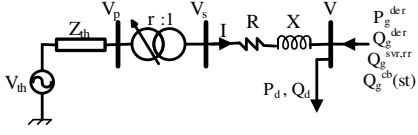}
  \caption{Simple example to showcase the proposed expediting technique.}\label{simple_system}
\end{figure}

Here, the characteristics that allow application of the expediting technique are shown through a simple example (Fig. \ref{simple_system}).
The control devices are all connected to the load point and their effects are aggregated in $\Delta P_g$ and $\Delta Q_g$ with $\rho$ as their energy price.
The transformer series impedance is included in the line impedance, $Z_{th}$=0, $X_M$ is neglected, $\tau=1$ h and $S^{base}$=1 MW. The approximate system cost is given in \eqref{Cost_Pg1} with $\rho_R$=0, $Q_d$=$Q_g$=0 and $|V|\approx 1$. Therefore, $|I| \approx P_d-P_g $. The cost in \eqref{Cost_Pg1} includes the \textsc{der}, upstream power purchase, copper loss and iron loss costs.
The only term that is neglected is $R(\Delta P_g)^2$.
This term is always positive. This is the key reasoning behind applying the expediting technique.

Mathematically speaking, the cost curve is concave up, i.e., is convex, since the second derivative (with respect to $P_g$) is always positive ($F''=2R\rho_A$). This does not mean the \textsc{dsp} is convex. It means that at any desired point the value of the simplified objective function is lower than the accurate cost.
\begin{equation}
\label{Cost_Pg1}
\begin{array}{l}
\displaystyle
F\approx \rho P_g + \rho_A \left(P_d-P_g+R(P_d-P_g)^2+\frac{{V_{th}}^2}{r^2R^n_c}\right)
\end{array}
\hspace{-3mm}
\end{equation}

\begin{figure}[!t]
\centering
\subfloat[As a function of $P_g^{der}$]{\includegraphics[width=0.467\columnwidth,trim=0.0cm 0.035cm 0.0cm 0.035cm,clip=true]{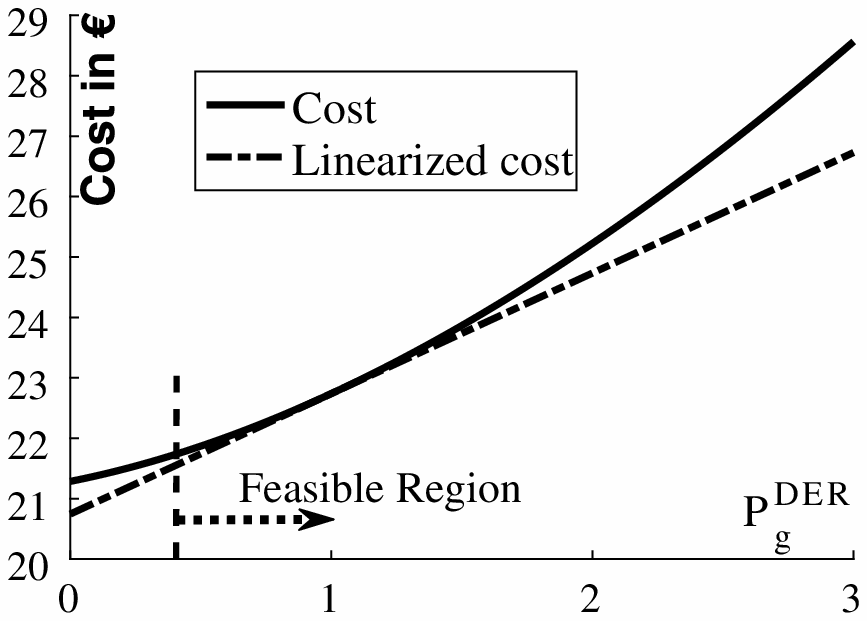}
\label{Simple_Example2}}
\subfloat[As a function of tap position]{\includegraphics[width=0.493\columnwidth,trim=-0.25cm 0.035cm 0.040cm 0.030cm,clip=true]{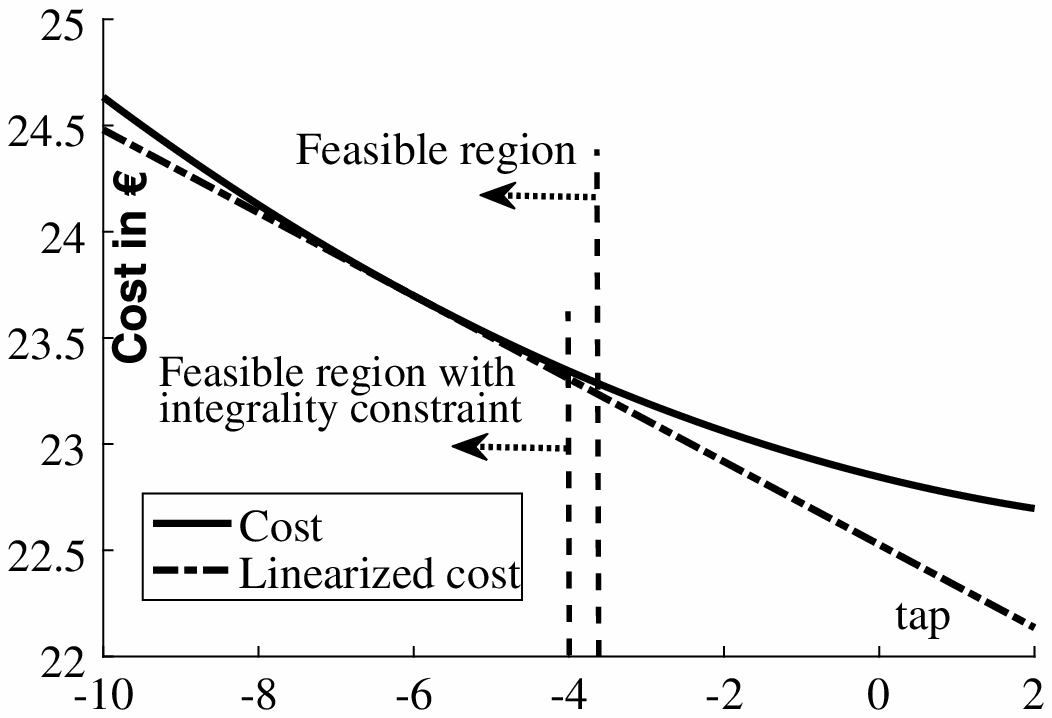}
\label{Simple_Example3}}
\caption{Accurate and linearized cost.}
\label{SIEx00}
\end{figure}

Fig. \ref{Simple_Example2} gives the accurate cost curve and linearized cost for a more realistic situation with $P_d$=1.5 MW, $Q_d$=0.5 MW, $R$=$X$=3 pu, $S^{base}$=100 MW, $V^{base}$=12.66 kV, $\tau$=0.25 h, $tap$=0, $\rho_A$=50 \euro/MWh, $\rho_R$=10 \euro/MVarh, $\rho$=60 \euro/MWh and $\hat{P}_g$=1 MW. The transformer data is presented in Table \ref{Transformers} for transformer 1 and load coefficients can be found in \cite{Nouri2017} for residential loads. The minimum and maximum allowable voltages are 0.95 and 1.05 pu., respectively. The cost curve is concave up and the linearized cost is always lower than the accurate cost.

The situation is the same for control variables $Q_g^{der}$, $Q_g^{svr}$, $Q_g^{rr}$ and $st$. Generally speaking, the reason lies in the fact that the active and reactive power losses of a distribution network can be expressed with a polynomial of degree 2 of these control variables with positive coefficients for the square terms, signifying an always positive second partial derivative for the system cost with respect to each control variable in this list.

To complete the discussion, the tap positions should also be taken into account. According to sub-section \ref{SecA}, increasing the tap position reduces the transformers' core loss.
However, this usually increases the line losses slightly.
With a higher tap position, the load point voltages and the active and reactive power demands are lower. The value of this reduction depends on the load types. More details can be found in \cite{Pouladi2019}. Fig. \ref{Simple_Example3} shows the accurate cost curve and the linearized cost as a function of the tap position. Other controllable devices are neglected. The cost curve is concave up and therefore, the linearized cost is always lower than the accurate cost. This can be generalized based on the results reported in \cite{Pouladi2019}. For the sake of brevity, the transformer core loss is selected here as one of the main components of the system loss which varies widely as the tap position is changed. Transformer core loss ($P_c$) and its partial derivatives are presented in \eqref{Core_loss}. The second derivative is always positive.
\begin{equation}
\label{Core_loss}
\begin{array}{l}
\displaystyle
P_c\approx\frac{{V_{th}}^2}{r^2R_c^n}, \hspace{3mm}
\frac{\partial{P_c}}{\partial{r}}\approx\frac{-2{V_{th}}^2}{r^3R_c^n}, \hspace{3mm}
\frac{\partial^2{P_c}}{{\partial{r}}^2}\approx\frac{6{V_{th}}^2}{r^4R_c^n} 
\end{array}
\end{equation}

In an optimization problem, if the second partial derivatives of the objective function with respect to the optimization variables are always non-negative over a certain region ($\Omega$), the linearized objective function ($\widetilde{F}$), i.e., the function obtained by linearizing the problem around any desired point ($X^0$) in this region, is lower than or equal to the accurate objective function ($F$) at any desired point ($X^*$) in this region \eqref{Linear_VS_Accurate}.
\begin{equation}
\label{Linear_VS_Accurate}
\begin{array}{l}
\displaystyle
\left[X^*,X^0\in{\Omega}\right] \wedge
\left[\frac{\partial^2{F}}{\partial{x_i^2}}\geq{0}
\hspace{3mm}
\forall x_i \in X, X\in{\Omega}\right]
\\
\displaystyle
\rightarrow
\widetilde{F}(X^*)
=
\sum^{N_{\text{W}}}_{i=1}{\left[\frac{\partial{F}}{\partial{x_i}}\right]^0
\Delta{x_i}}
+
F(X^0) 
\leq
F(X^*)
\end{array}
\end{equation}
\section{Case Studies}\label{Case_Studies}
IEEE 33-bus distribution test system \cite{Pouladi2019} is used to analyze the effectiveness of the proposed framework.
Two parallel transformers under \textsc{oltc} control (Table \ref{Transformers}) connect this system to the upstream system. A 1000 kVar 10-step \textsc{cb}, a 1200 kVA \textsc{der} and a 1500 kVar \textsc{svr} are added at buses 33, 14 and 30, respectively. The maximum power angle of the \textsc{der} is $30^\textsc{o}$.
\subsection{Analyzing the Objective Function and Constraints}\label{CaseA}
A step change has been applied to the voltage level of the upstream system and
the optimal solution is tracked.
The upstream model is updated based on subsection \ref{SecA}.

%
%
\newcolumntype{N}[1]{>{\centering\let\newline\\\arraybackslash\hspace{0pt}}m{#1}}
\newcolumntype{C}[1]{>{\centering\let\newline\\\arraybackslash\hspace{0pt}}m{#1}}
\newcolumntype{X}[1]{>{\centering\let\newline\\\arraybackslash\hspace{-2pt}}m{#1}}
\newcolumntype{R}[1]{>{\centering\let\newline\\\arraybackslash\hspace{-2pt}}m{#1}}
\newcolumntype{M}[1]{>{\centering\let\newline\\\arraybackslash\hspace{0pt}}m{#1}}
\newcolumntype{I}[1]{>{\centering\let\newline\\\arraybackslash\hspace{0pt}}m{#1}}
\newcolumntype{H}[1]{>{\centering\let\newline\\\arraybackslash\hspace{-2pt}}m{#1}}
\newcolumntype{L}[1]{>{\centering\let\newline\\\arraybackslash\hspace{-2pt}}m{#1}}
\newcolumntype{U}[1]{>{\centering\let\newline\\\arraybackslash\hspace{0pt}}m{#1}}
\begin{table}
    \caption{Transformers' Data}
		\label{Transformers}
		\centering
		\setlength\tabcolsep{5pt} 
		\setlength{\doublerulesep}{2\arrayrulewidth}
\scalebox{0.98}{
\begin{tabular}{ c | c | c | c | c | c | c | c | c}
\hline
& Cap.  & X  & R  & $X_M$  & $R_c$  & $tap^{max}$ & $tap^{min}$ & $\Delta{U}$ \\
    & (MVA) & (pu.) & (pu.) & (pu.) & (pu.) &   &  & (\%) \\
\hline
\hline
1 & 3 & 0.100  & 0.006 & 390 & 400 & 10 & -10 & 1 \\
\hline
2 & 3 & 0.110  & 0.006 & 380 & 400 & 12 & -12 & 1 \\
\hline
\end{tabular}}
\end{table}
In the first study, it is assumed that the \textsc{cb} step, transformers' taps and \textsc{der} reactive power are fixed at 0.
This allows to demonstrate the objective function and constraints' behaviors using three-dimensional figures.
Prices $\rho_A$, $\rho_{der}$ and $\rho_R$ are 50 \euro/MWh, 60 \euro/MWh and 10 \euro/MVarh, respectively
and $Z_{th}$=0.02+0.1$j$ pu. The initial Thevenin voltage and the voltage step change are 0.98 and +0.02, respectively.

\begin{figure}[!t]
  \centering
  \includegraphics[width=0.95\columnwidth,trim=1cm 0.92cm 0.5cm 0.91cm,clip=true]{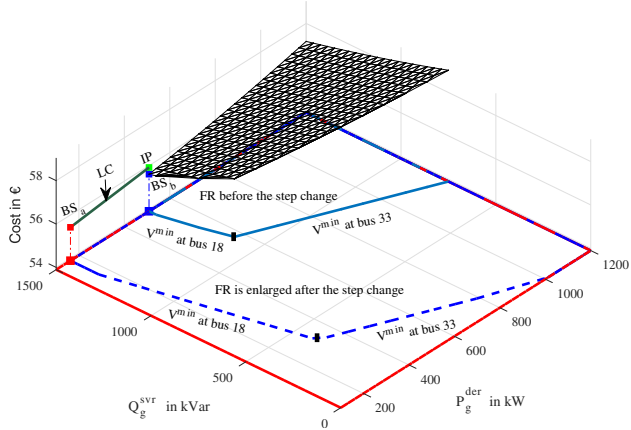}
	\caption{Operating cost as a function of $P_{g}^{der}$ and $Q_{g}^{svr}$ before the upstream voltage change and linearized model.}
\label{CostPdgQsvr}
\end{figure}

Fig. \ref{CostPdgQsvr} shows the value of objective function for the different values of control variables before the step change. All the constraints
have been mapped to the domain of control variables and the most restricting constraints have been found to form the feasible region (\textsc{fr}) before and after the step change in Fig. \ref{CostPdgQsvr}.
It is important to note that the cost function is concave up.

The exact solution of the \textsc{dsp} before the step change is given by $\text{BS}_\text{b}$. After applying the change, the feasible region is enlarged and though the current solution is still feasible, the operation cost is no longer optimal (point IP in Fig. \ref{CostPdgQsvr}). IP is also the starting point in the linearized formulation. The line LC shows the linearized objective function around the point IP (after the voltage step change), on $Q_{SVR}$=1500 kVar plane. The exact solution after the step change is shown by $\text{BS}_\text{a}$. The linearization error is quite low, since LC almost intercepts $\text{BS}_\text{a}$. 

\begin{figure}[!t]
  \centering
  \includegraphics[width=0.95\columnwidth,trim=1cm 0.90cm 0.5cm 0.99cm,clip=true]{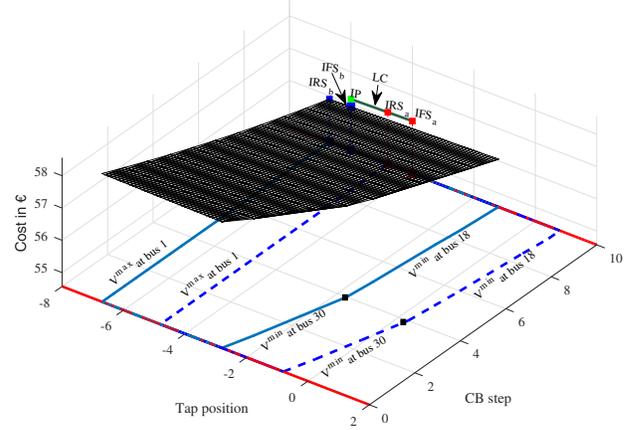}
	\caption{Operating cost as a function of $tap$ and $st$ before the change and linearized model.}\label{CostTapStep}
\end{figure}

In the second study, it is assumed that $P_{DER}$=500 kW and $Q_{SVR}$=500 kVar and $tap$ and $st$ can be changed to achieve the optimal solution. Fig. \ref{CostTapStep} shows the objective function for different values of these control variables before the voltage step change. The \textsc{fr} is also presented before and after the voltage step change. It should be noted that $tap$ and $st$ are integer variables. Therefore, the integer relaxed solution (\textsc{irs}) is not acceptable. The integer feasible solutions $\text{IFS}_\text{b}$ (before step change) and $\text{IFS}_\text{a}$ (after step change) are also shown. With the same set-point as $\text{IFS}_\text{b}$ the IP is achieved after the step change. With an acceptable accuracy, $\text{IRS}_\text{a}$ and $\text{IFS}_\text{a}$ are on the line LC, (the linearized objective function around IP). More importantly, the cost is concave up again.
\subsection{Scheduling Results}\label{CaseB}
The proposed framework is applied to solve the hourly \textsc{dsp}s in a 24-hour period. This paper tackles the short term \textsc{dsp} and therefore, for each hour a separate problem is solved.
The hourly load data and energy prices can be found in \cite{Pouladi2019} and $\rho_R=0.1\rho_A$. The load coefficients are provided in \cite{Nouri2017} for residential loads. Different studies are conducted.

\begin{figure}[t!]
  \centering
\includegraphics[width=0.91\columnwidth,trim=10 5 10 10,clip=true]{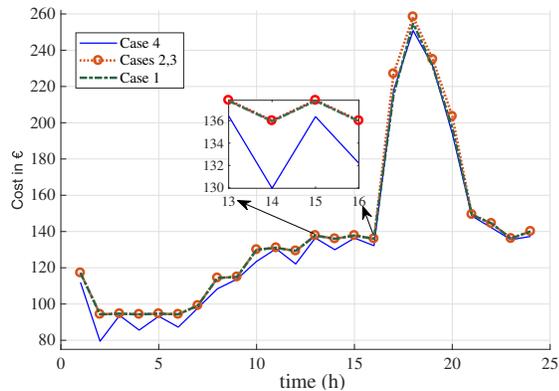}
  \caption{\textcolor[rgb]{0,0,0}{Operation costs in Cases 1-4.}}
\label{CPL0}
\end{figure}

In \textbf{Case 1}, the \textsc{dsp}s are solved with \textsc{lp} as the only sub-algorithm. It is assumed that at the starting point (\textsc{sp}), i.e., hour 0, the system schedule is optimal. Fig. \ref{CPL0} presents the results.
For simulation purpose, a power flow algorithm is used to find the actual voltages and currents and to build a more precise linearized model for the next hour. In a real system, this voltage and currents are being measured or estimated and the linearized model will be updated accordingly.

In \textbf{Case 2} the results are obtained by the proposed method based on \textsc{tra} and using the results of the simplified problem just to expedite the solution. These results are also presented in Fig. \ref{CPL0}. As can be seen, the quality of the solutions obtained with only \textsc{lp} as the sub-algorithm is acceptable except for hours with high variation in the system load (which in turn, require high variation in the set-points of the controllable devices).


In \textbf{Case 3}, $tap$, $st$, $P_{g}^{der}$ and $Q_{g}^{svr}$ are initially set to zero. This is a non-optimal and infeasible solution for hour 1. The results of the proposed method based on switching the sub-algorithm are the same as those obtained in Case 2. This means the proposed method is robust against the non-optimal \textsc{sp}.

In \textbf{Case 4}, the \textsc{dsp}s are solved beginning from the same \textsc{sp} using \textsc{lp}. In Case 4, the linearization error are higher compared to Case 1 in all hours as can be seen in Fig. \ref{CPL0}. At the first glance, it might seem as if at some periods the cost obtained in Case 4 are the same as those found in Case 1. However, as can be seen in the magnified part of Fig. \ref{CPL0}, this is not true.


In Case 1, the solutions found using \textsc{lp} were always feasible. In Case 4, ten infeasible solutions were found. Most infeasible solutions have happened during the light load periods. During these periods, there is no under-voltage issue, so \textsc{lp} tries to reduce the voltages as much as possible to reduce the system cost according to the voltage-dependent characteristics of the system loads. This leads to wrong tap-changing operations and infeasible solutions. \textsc{lp} tries to push the solution towards the FR for the next hour. Therefore, the operation cost fluctuates in this case.
It is fascinating that the linearized objective function is always lower than the accurate objective function. This validates the explanation provided in subsection \ref{IE}.

The solution times depend on the \textsc{sp}s and vary for different hours, signifying the case dependent nature of \textsc{bc} algorithm.
In Case 1, 2 and 3, the average solution times are 463, 705 and 1933 ms, respectively.
However, the solution optimality cannot be guaranteed in Case 1.
\begin{figure}[!t]
  \centering
  \includegraphics[width=0.73\columnwidth,clip=true]{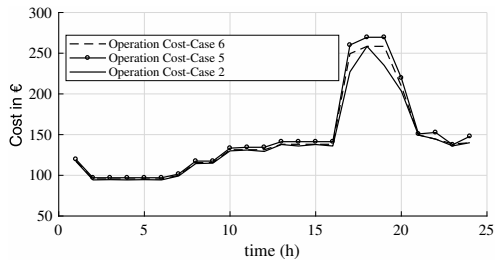}
	\caption{Cost for Cases 2, 5 and 6.}\label{CCVR0}
\end{figure}

\begin{figure}[!tb]
  \centering
  \includegraphics[width=0.88\columnwidth,trim=1.10cm 1.32cm 1.05cm 0.30cm,clip=true]{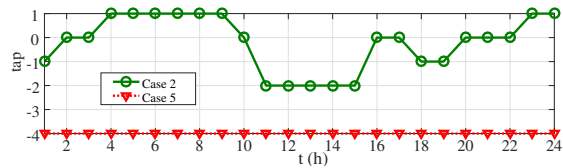}
	\caption{Tap positions in Case 2 and Case 5.}\label{TapPos0}
\end{figure}

\textbf{Case 5} is designed to show the effects of neglecting the voltage dependent nature of loads.
Since the loads are not really constant power, after the optimization is converged, the real costs are calculated using a power flow considering the voltage-dependent nature of the loads. Fig. \ref{CCVR0} shows these costs as well as the costs obtained assuming that the loads are really constant power (\textbf{Case 6}). Comparing the results of Case 6 and Case 2 shows the effects of load characteristics. The operation costs are higher for voltage independent loads (Case 6).
\begin{figure}[!tb]
  \centering
  \includegraphics[width=0.72\columnwidth,trim=9 52 10 27,clip=true]{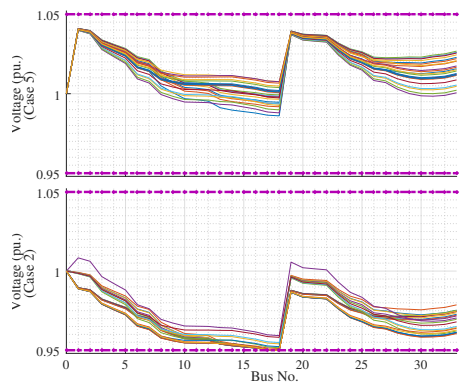}
	\caption{Voltages profiles in 24 hours for Case 2 and Case 5.}\label{VoltProf}
\end{figure}
\begin{table}[!tb]
    \caption{Energy Demand and Energy Loss}
		\label{CVREff}
		\centering
		\setlength\tabcolsep{5pt} 
		\setlength{\doublerulesep}{2\arrayrulewidth}
\scalebox{1.00}{
\begin{tabular}{X{3.33cm} | N{1.75cm} | C{1.17cm} | M{0.98cm} }
\hline
Strategy & Total Energy Demand (kWh) & Total Loss (kWh) & Total Cost (\euro) \\
\hline\hline
Voltage dependent load model & 44140.02 & 7527.33 & 3334.23 \\
\hline
Constant power load model & 44700.24 & 7850.23 & 3516.84 \\
\hline
\end{tabular}}
\end{table}
Comparing the results of Case 2 and Case 5 will show the effects of load model accuracy.
The optimal operation costs are also presented in Fig. \ref{CCVR0} for Case 2. The operation costs are much higher if voltage dependent nature of loads is not taken into account (Case 5).
Table \ref{CVREff} gives the total costs, energy consumptions and energy losses in Cases 2 and 5.
The total energy loss is also lower for Case 2. The voltage profiles at all 24 hours are presented in Fig. \ref{VoltProf}. As can be seen with voltage dependent load model, the scheduling algorithm tries to keep the voltages as low as possible to reduce the active and reactive power demands. Fig. \ref{TapPos0} presents the tap position in cases 2 and 5. In Case 5, the tap position is set to -4 for all hours to keep the voltages as high as possible to reduce the copper losses. Lower tap positions lead to secondary side over-voltage. In Case 2, during the low-load periods, the tap position is increased to reduce the voltages as much as possible.
\begin{table*}[!t]
    \caption{\textcolor[rgb]{0,0,0}{Comparison between the proposed method and \textsc{socp} for solving the \textsc{dsp}}}
		\addtolength{\tabcolsep}{-3pt}
		\label{Compar}
		\centering
		\setlength\tabcolsep{5pt} 
		\setlength{\doublerulesep}{2\arrayrulewidth}

\begin{tabular}{>{\color{black}}X{1.7cm} | >{\color{black}}N{7cm} | >{\color{black}}C{8.30cm} }
\hline
  & Proposed method & \textsc{socp} based on branch flow technique  \\
\hline\hline
\textsc{dcd}s        & Can easily model the \textsc{dcd}s with desired level of accuracy. & Difficulty with \textsc{dcd}s. Needs extra integer variables or complicating approaches with simplifying assumptions such as those proposed by \cite{Shukla2019}. \\
\hline
Convergence          & Globally convergent algorithm (see Section \ref{SecIIIC}) & High quality solutions \cite{Shukla2019} under simplifying assumptions  \\
\hline
Speed    & Fast when combined with the proposed expediting technique & Fast due to the availability of commercial solvers for mixed integer \textsc{socp}  \\
\hline
Constraint handling  & Able to handle any nonlinear constraints & Constraints should be the linear functions of $\left|V_b\right|^2$ and $\left|I_l\right|^2$ or can be converted to convex cones.\\
\hline
Unbalanced systems   & Can be applied on practical unbalanced systems. & Simplifying assumptions should be made like those applied in \cite{Unbalanced} and \cite{UnbalancedAlaki}. These assumption are not acceptable for practical systems.  \\
\hline
Load model           & Any nonlinear voltage-dependent load model & Active/reactive loads should be expressed as the functions of $\left|V_b\right|^2$. \\
\hline
Single-phase control devices      & Controllable devices can be three phase, independent-per-phase or single phase.  & Cannot be applied with independent per-phase devices. \\
\hline
\end{tabular}
\end{table*}
\subsection{\textcolor[rgb]{0,0,0}{Comparison with Other Fast Scheduling Techniques}}\label{CaseC}
\textcolor[rgb]{0,0,0}{The efficiency of proposed method is compared with the state-of-the-art \textsc{socp} based on branch flow technique \cite{Stoch}.
In branch flow model, the network equations are rewritten in terms of $\left|V_b\right|^2$ and $\left|I_l\right|^2$.
These terms are replaced by linear variables and the branch flow equality constraints are replace by their conic programming inequality counterparts.
Further details are found in \cite{Stoch}. This convex relaxation is exact if the network topology is radial and the objective function is non-decreasing as the loads increase.
The main advantage of \textsc{socp} technique is the availability of commercial solvers.
Under some assumptions, the method was proved to be efficient for the balanced radial systems.
As discussed in Section \ref{sec_1}, the restricting assumptions that should be made to attain the exact convex relaxation render this technique inapplicable on most practical systems.
Table \ref{Compar} summarizes the comparison between the proposed and \textsc{socp} methods based on subsection \ref{LitRev} and the results of a case study which come next.}
\begin{figure}[!t]
  \centering
  \includegraphics[width=0.80\columnwidth,trim=0.2cm 0.00cm 1.0cm 0.30cm,clip=true]{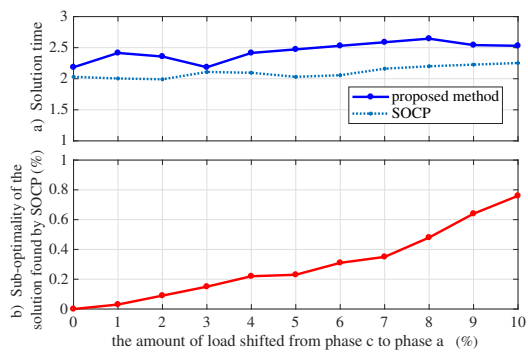}
	\caption{\textcolor[rgb]{0,0,0}{Comparison between the proposed method and \textsc{socp}.}}\label{ComparFig}
\end{figure}

\textcolor[rgb]{0,0,0}{To better demonstrate the points made in Table \ref{Compar}, 10\% of the load on phase $c$ of the IEEE 33-bus test system is shifted in 10 steps to phase $a$ and the \textsc{dsp} is solved by both proposed method and \textsc{socp}. The peak load level (hour 6 PM) is considered. It is assumed that all schedulable devices are independent-per-phase controllable devices.
To apply \textsc{socp} in unbalanced conditions, the simplifying assumption and the method proposed in \cite{Unbalanced} are used and the resultant problem is solved using CPLEX in GAMS. Fig. \ref{ComparFig} shows the solution times.
The solution times are quite acceptable for the proposed method, but slightly higher than the solution times with \textsc{socp}.
The reason mostly lies in the fact that the \textsc{socp} problems are fully solved with a commercial solver.
The expediting \textsc{lp} problems and quadratic \textsc{tra} sub-problems are also solved by a commercial solver.
On the other hand, the proposed method outperforms \textsc{socp} in terms of optimality. As the loads become more unbalanced, the solutions of \textsc{socp} become more sub-optimal. The reason is the approximate formulation used to attain the \textsc{socp} convex relaxed formulation. The proposed method seeks the solutions for independent-per-phase voltage control which allows lower voltages on phase $c$ compared to the solutions found using \textsc{socp}. These lower voltages reduce the load level and power loss on this phase.}
\section{Conclusions}\label{conclusion}
The results validate the proposed fast scheduling framework. The consideration of the voltage dependent nature of loads efficiently reduces the system cost. The expected energy saving may not be realized, if the voltage-dependent load model is not kept updated.
\textcolor[rgb]{0,0,0}{Compared to the available fast scheduling approaches, the proposed method can easily be applied on the practical unbalanced systems with any nonlinear voltage-dependent load models. This method is able to handle any kind of non-linear constraints.
With \textsc{lp} as the only sub-algorithm, the results are sub-optimal or even infeasible.
However, it was shown that using the solution of the simplified problem based on \textsc{lp} and expediting techniques, a fast decision can be made on stopping most of the branches during the branching process of \textsc{bc}.
The techniques provided in Section \ref{SecIIIC} help to faster obtain an upper bound on the value of the objective function of \textsc{minlp} problem. This further expedites the solution without compromising the optimality.} 
Using the proposed \textsc{tra}-based \textsc{minlp} method and novel expediting techniques, the quality of solution is guaranteed for the near-real-time applications.

\textcolor[rgb]{0,0,0}{The proposed formulation can be extended to include the uncertainties associated with \textsc{rr}s in future studies. There are other \textsc{tra}s and non-linear optimization algorithms that can be used instead of the \textsc{tra} applied in this study. Analyzing the performance of these algorithms for solving the \textsc{dsp} is also proposed for future research activities on this topic.}

\bibliographystyle{IEEEtran}
\bibliography{ref}

\begin{thebibliography}{10}
\providecommand{\url}[1]{#1}
\csname url@samestyle\endcsname
\providecommand{\newblock}{\relax}
\providecommand{\bibinfo}[2]{#2}
\providecommand{\BIBentrySTDinterwordspacing}{\spaceskip=0pt\relax}
\providecommand{\BIBentryALTinterwordstretchfactor}{4}
\providecommand{\BIBentryALTinterwordspacing}{\spaceskip=\fontdimen2\font plus
\BIBentryALTinterwordstretchfactor\fontdimen3\font minus
  \fontdimen4\font\relax}
\providecommand{\BIBforeignlanguage}[2]{{%
\expandafter\ifx\csname l@#1\endcsname\relax
\typeout{** WARNING: IEEEtran.bst: No hyphenation pattern has been}%
\typeout{** loaded for the language `#1'. Using the pattern for}%
\typeout{** the default language instead.}%
\else
\language=\csname l@#1\endcsname
\fi
#2}}
\providecommand{\BIBdecl}{\relax}
\BIBdecl

\bibitem{Watson2018}
J.~D. {Watson}, N.~R. {Watson}, and I.~{Lestas}, ``Optimized dispatch of energy
  storage systems in unbalanced distribution networks,'' \emph{IEEE Trans.
  Sustain. Energy}, vol.~9, no.~2, pp. 639--650, April 2018.

\bibitem{KIANMEHR2019471}
E.~kianmehr, S.~Nikkhah, and A.~Rabiee, ``Multi-objective stochastic model for
  joint optimal allocation of dg units and network reconfiguration from dg
  owner’s and disco’s perspectives,'' \emph{Renewable Energy}, vol. 132,
  pp. 471 -- 485, 2019.

\bibitem{Songqiang2018}
S.~Qiu and Z.~Chen, ``An interior point method for nonlinear optimization with
  a quasi-tangential subproblem,'' \emph{J. Comput. Appl. Math.}, vol. 334, pp.
  77 -- 96, 2018\color{black}.

\bibitem{Sheng2014}
W.~{Sheng}, K.~{Liu}, and S.~{Cheng}, ``Optimal power flow algorithm and
  analysis in distribution system considering distributed generation,''
  \emph{IET Gener. Transm. Distrib.}, vol.~8, no.~2, pp. 261--272, February
  2014.

\bibitem{Dao2017}
T.~V. Dao, S.~Chaitusaney, and H.~T.~N. Nguyen, ``Linear least-squares method
  for conservation voltage reduction in distribution systems with photovoltaic
  inverters,'' \emph{IEEE Trans. Smart Grid}, vol.~8, no.~3, pp. 1252--1263,
  May 2017.

\bibitem{BAHRAMI2020}
S.~Bahrami and K.~Amini, ``An efficient two-step trust-region algorithm for
  exactly determined consistent systems of nonlinear equations,'' \emph{J.
  Comput. Appl. Math.}, vol. 367, 2020\color{black}.

\bibitem{Stoch}
H.~{Gao}, J.~{Liu}, and L.~{Wang}, ``Robust coordinated optimization of active
  and reactive power in active distribution systems,'' \emph{IEEE Trans. Smart
  Grid}, vol.~9, no.~5, pp. 4436--4447, Sep. 2018.

\bibitem{LinearOLTC}
W.~{Wu}, Z.~{Tian}, and B.~{Zhang}, ``An exact linearization method for oltc of
  transformer in branch flow model,'' \emph{IEEE Trans. Power Syst.}, vol.~32,
  no.~3, pp. 2475--2476, May 2017.

\bibitem{Unbalanced}
B.~A. {Robbins} and A.~D. {Domínguez-García}, ``Optimal reactive power
  dispatch for voltage regulation in unbalanced distribution systems,''
  \emph{IEEE Trans. Power Syst.}, vol.~31, no.~4, pp. 2903--2913, July 2016.

\bibitem{UnbalancedAlaki}
Y.~{Gu}, H.~{Jiang}, J.~J. {Zhang}, Y.~{Zhang}, H.~{Wu}, and E.~{Muljadi},
  ``Multi-timescale three-phase unbalanced distribution system operation with
  variable renewable generations,'' \emph{IEEE Transactions on Smart Grid},
  vol.~10, no.~4, pp. 4497--4507, July 2019.

\bibitem{Arefifar2013}
S.~A. Arefifar and W.~Xu, ``Online tracking of voltage-dependent load
  parameters using ultc created disturbances,'' \emph{IEEE Trans. Power Syst.},
  vol.~28, no.~1, pp. 130--139, Feb 2013.

\bibitem{Bahadornejad2014}
M.~Bahadornejad and N.~K.~C. Nair, ``Intelligent control of on-load tap
  changing transformer,'' \emph{IEEE Trans. Smart Grid}, vol.~5, no.~5, pp.
  2255--2263, Sept 2014.

\bibitem{Ren2018}
J.~{Ren}, J.~{Hu}, R.~{Deng}, D.~{Zhang}, Y.~{Zhang}, and X.~{Shen}, ``Joint
  load scheduling and voltage regulation in the distribution system with
  renewable generators,'' \emph{IEEE Trans Ind. Informat.}, vol.~14, no.~4, pp.
  1564--1574, April 2018.

\bibitem{Li2016}
J.~Li, H.~Xin, W.~Wei, and W.~Dai, ``Decentralised conic optimisation of
  reactive power considering uncertainty of renewable energy sources,''
  \emph{IET Renew. Power Gen.}, vol.~10, no.~9, pp. 1348--1355, 2016.

\bibitem{Pouladi2019}
A.~{Pouladi}, A.~K. {Zadeh}, and A.~{Nouri}, ``Control of parallel ultc
  transformers in active distribution systems,'' \emph{IEEE Syst. J.}, to be
  published\color{black}.

\bibitem{Nouri2017}
M.~{Gheydi}, A.~{Nouri}, and N.~{Ghadimi}, ``Planning in microgrids with
  conservation of voltage reduction,'' \emph{IEEE Syst. J.}, vol.~12, no.~3,
  pp. 2782--2790, Sep. 2018.

\bibitem{Dey2018}
\BIBentryALTinterwordspacing
S.~S. Dey and M.~Molinaro, ``Theoretical challenges towards cutting-plane
  selection,'' \emph{Mathematical Programming}, vol. 170, no.~1, pp. 237--266,
  Jul 2018. [Online]. Available:
  \url{https://doi.org/10.1007/s10107-018-1302-4}
\BIBentrySTDinterwordspacing

\bibitem{Shukla2019}
S.~R. {Shukla}, S.~{Paudyal}, and M.~R. {Almassalkhi}, ``Efficient distribution
  system optimal power flow with discrete control of load tap changers,''
  \emph{IEEE Trans. Power Syst.}, vol.~34, no.~4, pp. 2970--2979, July 2019.

\end{thebibliography}

\begin{IEEEbiographynophoto}{Alireza Nouri}
(M’17) received the Ph.D. degree in electrical engineering from the Sharif University of Technology, Tehran, Iran, in 2016. He is now a senior power system researcher with the School of Electrical and Electronic Engineering, University College Dublin. His current research has been focused on power systems optimization and control.
\end{IEEEbiographynophoto}

\begin{IEEEbiographynophoto}{Alireza Soroudi}
(M’13–SM’16) received the Ph.D. degree in electrical engineering from the Grenoble-INP, Grenoble, France, in 2012. He is an Assistant Professor at UCD. His research interests include power systems planning and operation, risk, and uncertainty modeling.
\end{IEEEbiographynophoto}

\begin{IEEEbiographynophoto}{Andrew Keane}
(S’04–M’07–SM’14) received the Ph.D. degree in electrical engineering from the University College Dublin (UCD), Dublin, Ireland, in 2007. He is a Professor and Director of the Energy Institute at UCD. His research interests include power systems planning and operation, distributed energy resources, and distribution networks.
\end{IEEEbiographynophoto}
\end{document}